\pdfoutput=1  
\documentclass[sigconf]{acmart}

\makeatletter                   
\def\mdseries@tt{m}             
\makeatother                    

\usepackage[plain]{fancyref}
\usepackage[draft=true]{minted} 
\usepackage{color}
\usepackage{textgreek}
\usepackage{breakurl}           

\usepackage{amssymb}
\usepackage{lipsum}
\usepackage[final]{pdfpages}
\usepackage{setspace}
\usepackage{epsf}
\usepackage{epsfig}
\usepackage{graphicx}
\usepackage{algorithmic}
\usepackage{amsmath}
\usepackage{letltxmacro}

\usepackage{enumitem}
\usepackage{tikz}
\usepackage{xspace}

\usepackage{url}
\usepackage{multirow}

  \renewenvironment{thebibliography}[1]{%
    \begin{oldthebibliography}{#1}%
      \setlength{\parskip}{0ex}%
      \setlength{\itemsep}{0ex}%
  }%
  {%
    \end{oldthebibliography}%
  }

\newwrite\arxivdeps
\immediate\openout\arxivdeps=\jobname-arxivdeps.log

\newcommand\verifymarkedforarxivfile[1]{%
\ifdefined\arxivbuild
\else
\IfFileExists{#1}%
{}%
{\GenericWarning{Marked file (#1) for inclusion in arxiv build does not exist}}
\fi%
}

\newcommand\markforarxiv[1]{%
\verifymarkedforarxivfile{#1}%
\write\arxivdeps{IncludeInArxiv: #1}%
}

\DeclareUrlCommand\UScore{\urlstyle{rm}}

\LetLtxMacro\oldincludegraphics\includegraphics
\renewcommand{\includegraphics}[2][]{%
\markforarxiv{#2}%
\oldincludegraphics[#1]{#2}}

\LetLtxMacro\oldincludepdf\includepdf
\renewcommand{\includepdf}[2][]{%
\markforarxiv{#2}%
\oldincludepdf[#1]{#2}}

\def\nfigure[#1,#2,#3]{
\begin{figure}
\vspace*{0mm}
\begin{center}

\includegraphics[width=\columnwidth]{#1} 
\vspace*{-6mm}\caption[]{#2
} \label{#3}

\vspace*{-3mm}
\end{center}
\end{figure}}

\def\cfigure[#1,#2,#3]{
\begin{figure}
\vspace*{0mm}
\begin{center}

\includegraphics[width=3in]{#1} 
\vspace*{-3mm}\caption[]{#2
} \label{#3}
 
\vspace*{-5mm}
\end{center}
\end{figure}}

\def\cfigurefour[#1,#2,#3]{
\begin{figure}
\vspace*{0mm}
\begin{center}

\includegraphics[width=4in]{#1} 
\vspace*{-3mm}\caption[]{#2
} \label{#3}
 
\vspace*{-5mm}
\end{center}
\end{figure}}

\def\cfiguretemp[#1,#2,#3]{
\begin{figure}
\vspace*{0mm}
\begin{center}

\includegraphics[width=3.5in]{#1} 
\vspace*{-3mm}\caption[]{#2
} \label{#3}
 
\vspace*{-5mm}
\end{center}
\vspace*{-2mm}
\end{figure}}

\def\wfigure[#1,#2,#3]{
\begin{figure*}
\vspace*{0mm}
\begin{center}
 \includegraphics[width=\textwidth]{#1} 
 \vspace*{-3mm}\caption[]{#2
} \label{#3}
 
\end{center}
\end{figure*}}

\def\threefigure[#1,#2,#3,#4,#5]{
\begin{figure*}
\vspace*{0mm}
\begin{center}

\begin{tabular}{ccc}
\includegraphics[width=2in]{#1} & \includegraphics[width=2in]{#2} &  \includegraphics[width=2in]{#3} \\
(a) & (b) & (c) \\
\end{tabular}
\vspace*{-3mm}\caption[]{#4
} \label{#5}

\vspace*{-5mm}
\end{center}
\vspace*{-2mm}
\end{figure*}}

\def\dcfigure[#1,#2,#3,#4,#5,#6]{
{
\begin{figure*}
\begin{center}
\begin{minipage}[c]{\columnwidth}{
\includegraphics[width=\columnwidth]{#1} 
\vspace*{0mm}\caption[]{#2} \label{#3} \
}\end{minipage}\hspace*{\columnsep}\
\begin{minipage}[c]{\columnwidth}{
\includegraphics[width=\columnwidth]{#4} 
\vspace*{0mm}\caption[]{#5}\label{#6} \
}\end{minipage}
\end{center}
\end{figure*}
}
}

\def\scfigure[#1,#2,#3]{
{
\begin{figure*}
\begin{center}
\begin{minipage}[c]{3.5in}{
\includegraphics[width=3.5in]{#1} 
}\end{minipage}
\caption[]{#2} \label{#3} \
\end{center}
\end{figure*}
}
}

\def\tableByTable[#1,#2,#3,#4,#5,#6]{
{
\begin{table*}
\begin{center}
\begin{minipage}[c]{3in}{
\centering
{#1}
\vspace*{0mm}\tabcaption[]{#2}\label{#3} \
}\end{minipage}\hspace*{\columnsep}\
\begin{minipage}[c]{3in}{
\centering
{#4}
\vspace*{0mm}\tabcaption[]{#5}\label{#6} \
}\end{minipage}
\end{center}
\end{table*}
}
}

\def\figureByTable[#1,#2,#3,#4,#5,#6]{
{
\begin{figure*}
\begin{center}
\begin{minipage}[c]{3in}{
\centering
\includegraphics[width=\textwidth]{#1}
\vspace*{0mm}\figcaption[]{#2} \label{#3} \
}\end{minipage}\hspace*{\columnsep}\
\begin{minipage}[c]{3.3in}{
\centering
{#4}
\vspace*{0mm}\tabcaption[]{#5}\label{#6} \
}\end{minipage}
\end{center}
\end{figure*}
}
}

\def\tableByFigure[#1,#2,#3,#4,#5,#6]{
{
\begin{figure*}
\begin{center}
\begin{minipage}[c]{4.3in}{
\centering
{#1}
\vspace*{0mm}\tabcaption[]{#2} \label{#3} \
}\end{minipage}\hspace*{\columnsep}\
\begin{minipage}[c]{2.2in}{
\centering
\includegraphics[width=\textwidth]{#4}
\vspace*{-0.35in}\caption[]{#5}\label{#6} \
}\end{minipage}
\end{center}
\end{figure*}
}
}

\def\doublecfigure[#1,#2,#3,#4]{
{
\begin{figure}
\begin{center}
\begin{minipage}[c]{1.5in}{
\begin{center}
\includegraphics[width=1.5in]{#1}
\end{center}
}\end{minipage}\hspace*{1em}\
\begin{minipage}[c]{1.5in}{
\begin{center}
\includegraphics[width=1.5in]{#2}
\end{center}
}\end{minipage}
\vspace*{0mm}\caption[]{#3} \label{#4} \
\end{center}
\end{figure}
}
}

\def\qcfigure[#1,#2,#3,#4,#5,#6]{
{
\begin{figure*}
\vspace*{0.2in}\
\begin{center}
\begin{minipage}[c]{3in}{
\includegraphics[width=3in]{#1} 
\vspace*{-3mm}
}
\end{minipage}\hspace*{0.5in}\
\begin{minipage}[c]{3in}{
\includegraphics[width=3in]{#2} 
\vspace*{-3mm}
}\end{minipage}

\begin{minipage}[c]{3in}{
\includegraphics[width=3in]{#3} 
\vspace*{-3mm}
}
\end{minipage}\hspace*{0.5in}\
\begin{minipage}[c]{3in}{
\includegraphics[width=3in]{#4} 
\vspace*{-3mm}
}\end{minipage}
\end{center}
\caption[]{#5}\label{#6}
\end{figure*}
}
}

\def\twfigure[#1,#2,#3,#4,#5]{
{
\begin{figure*}
\vspace*{0.2in}\
\begin{center}
\begin{minipage}[c]{6.5in}{
\includegraphics[width=6.5in]{#1} 
\vspace*{-3mm}
}
\end{minipage}

\begin{minipage}[c]{6.5in}{
\includegraphics[width=6.5in]{#2} 
\vspace*{-3mm}
}\end{minipage}

\begin{minipage}[c]{6.5in}{
\includegraphics[width=6.5in]{#3} 
\vspace*{-3mm}
}
\end{minipage}
\end{center}
\caption[]{#4}\label{#5}
\end{figure*}
}
}

\def\dwfigure[#1,#2,#3,#4]{
{
\begin{figure*}
\vspace*{0.2in}\
\begin{center}
\begin{minipage}[c]{6.5in}{
\includegraphics[width=6.5in]{#1} 
\vspace*{-3mm}
}
\end{minipage}

\begin{minipage}[c]{6.5in}{
\includegraphics[width=6.5in]{#2} 
\vspace*{-3mm}
}\end{minipage}

\end{center}
\caption[]{#3}\label{#4}
\end{figure*}
}
}

\def\dssfigure[#1,#2,#3,#4,#5,#6]{
{
\begin{figure*}
\vspace*{0.2in}\
\begin{center}
\begin{minipage}[c]{4in}{
\includegraphics[width=4in]{#1}
\vspace*{-3mm}\caption[]{#2} \label{#3} \
}\end{minipage}\hspace*{0.5in}\
\begin{minipage}[c]{2in}{
\includegraphics[width=2in]{#4}
\vspace*{-3mm}\caption[]{#5}\label{#6} \
}\end{minipage}
\end{center}
\vspace*{-0.4in}\
\end{figure*}
}
}

\def\dsfigure[#1,#2,#3,#4,#5,#6]{
{
\begin{figure*}
\vspace*{0.2in}\
\begin{center}
\begin{minipage}[c]{3in}{
\includegraphics[width=3in]{#1}
\vspace*{-3mm}\caption[]{#2} \label{#3} \
}\end{minipage}\hspace*{0.5in}\
\begin{minipage}[c]{3in}{
\hspace*{0.5in}\
\includegraphics[height=3in]{#4}
\vspace*{-3mm}\caption[]{#5}\label{#6} \
}\end{minipage}
\end{center}
\vspace*{-0.4in}\
\end{figure*}
}
}

\def\dsyfigure[#1,#2,#3,#4,#5,#6]{
{
\begin{figure*}
\vspace*{0.2in}\
\begin{center}
\begin{minipage}[c]{2.5in}{
\includegraphics[height=2.5in]{#1}
\vspace*{-3mm}\caption[]{#2} \label{#3} \
}\end{minipage}\hspace*{0.5in}\
\begin{minipage}[c]{2.5in}{
\includegraphics[height=2.5in]{#4}
\vspace*{-3mm}\caption[]{#5}\label{#6} \
}\end{minipage}
\end{center}
\vspace*{-0.4in}\
\end{figure*}
}
}

\def\dyfigure[#1,#2,#3,#4,#5,#6]{
{
\begin{figure*}
\vspace*{0.2in}\
\begin{center}
\begin{minipage}[c]{3in}{
\includegraphics[height=3in]{#1} 
\vspace*{-3mm}\caption[]{#2} \label{#3} \
}\end{minipage}\hspace*{0.5in}\
\begin{minipage}[c]{3in}{
\includegraphics[height=3in]{#4} 
\vspace*{-3mm}\caption[]{#5}\label{#6} \
}\end{minipage}
\end{center}
\vspace*{-0.4in}\
\end{figure*}
}
}

\def\dyoldfigure[#1,#2,#3,#4,#5,#6]{
{
\begin{figure*}
\vspace*{0.2in}\
\begin{center}
\begin{minipage}[c]{3in}{
\epsfysize=2.0in\
\hspace{0.5in}\
\epsfbox{#1}
\vspace*{-3mm}\caption[]{#2} \label{#3} \
}\end{minipage}\hspace*{0.25in}\
\begin{minipage}[c]{3in}{
\epsfysize=2.0in\
\hspace{0.5in}\
\epsfbox{#4}
\vspace*{-3mm}\caption[]{#5}\label{#6} \
}\end{minipage}
\end{center}
\vspace*{-0.4in}\
\end{figure*}
}
}

\def\cfiguredouble[#1,#2,#3,#4]{
\begin{figure}
\vspace*{0.2in}\
\begin{center}
\begin{minipage}[c]{1.5in}{
\epsfxsize=1.5in\
\epsfbox{#1}
}\end{minipage}\hspace*{0.1in}\
\begin{minipage}[c]{1.5in}{
\epsfxsize=1.5in\
\vspace{0.1in}\epsfbox{#2}
}\end{minipage}\vspace*{-0.10in} \caption[]{#3}\label{#4}
\end{center}
\vspace*{-0.4in}\
\end{figure}
}

\def\wpfigure[#1,#2,#3,#4]{
\begin{figure*}
\vspace*{4mm}
\begin{center}

\includegraphics[width=#4]{#1} 

\vspace*{-3mm}\caption[]{#2
} \label{#3}

\vspace*{-5mm}
\end{center}
\end{figure*}}

\def\wprfigure[#1,#2,#3,#4,#5]{
\begin{figure*}
\vspace*{4mm}
\begin{center}

\includegraphics[width=#4, angle=#5]{#1} 

\vspace*{-3mm}\caption[]{#2
} \label{#3}

\vspace*{-5mm}
\end{center}
\end{figure*}}

\def\DoubleFigureWSlide[#1,#2,#3,#4,#5,#6,#7,#8,#9]{
\begin{figure*}
\vspace*{#9}
\begin{center}
\begin{minipage}{#4}
\includegraphics[width=#4]{#1}
\vspace*{-3mm}\caption{#2
}\label{#3}
\end{minipage}
\hspace{2em}
\begin{minipage}{#8}
\includegraphics[width=#8]{#5}
\vspace*{-3mm}\caption{#6
}\label{#7}
\end{minipage}
\vspace*{-5mm}
\end{center}
\end{figure*}
}

\def\DoubleFigureW[#1,#2,#3,#4,#5,#6,#7,#8]{
\begin{figure*}
\vspace*{0in}
\begin{center}
\begin{minipage}{#4}
\includegraphics[width=#4]{#1}
\vspace*{-3mm}\caption{#2
}\label{#3}
\end{minipage}
\hspace{2em}
\begin{minipage}{#8}
\includegraphics[width=#8]{#5}
\vspace*{-3mm}\caption{#6
}\label{#7}
\end{minipage}
\vspace*{-5mm}
\end{center}
\end{figure*}
}

\def\DoubleFigureWHack[#1,#2,#3,#4,#5,#6,#7,#8]{
\begin{figure*}
\vspace*{0in}
\begin{center}
\begin{minipage}{3in}
\includegraphics[width=#4]{#1}
\vspace*{-3mm}\caption{#2
}\label{#3}
\end{minipage}
\hspace{2em}
\begin{minipage}{3in}
\includegraphics[width=#8]{#5}
\vspace*{-3mm}\caption{#6
}\label{#7}
\end{minipage}
\vspace*{-5mm}
\end{center}
\end{figure*}
}

\def\ddcfigure[#1,#2,#3,#4]{
\begin{figure*}
\vspace*{0.2in}\
\begin{center}
\begin{minipage}[c]{\columnwidth}{
\includegraphics[width=\columnwidth]{#1} 
}\end{minipage}\hspace{0.5in}\
\begin{minipage}[c]{\columnwidth}{
\includegraphics[width=\columnwidth]{#2} 
}\end{minipage} \caption[]{#3}\label{#4}
\end{center}
\end{figure*}
}

\def\ddcfigureSlide[#1,#2,#3,#4,#5]{
\begin{figure*}
\vspace*{#5}\
\begin{center}
\begin{minipage}[c]{3in}{
\includegraphics[height=3in]{#1} 
}\end{minipage}\hspace{0.5in}\
\begin{minipage}[c]{3in}{
\includegraphics[height=3in]{#2} 
}\end{minipage}\vspace*{-0.10in} \caption[]{#3}\label{#4}
\end{center}
\vspace*{-0.4in}\
\end{figure*}
}

\def\cxfigure[#1,#2,#3]{
\begin{figure}
\vspace*{4mm}
\begin{center}
 
\epsfxsize=2.5in\
\epsfbox{#1}\
 
\vspace*{-0.10in}\caption[]{#2
} \label{#3}
 
\vspace*{-5mm}
\end{center}
\vspace*{-2mm}
\end{figure}}

\newif\ifremark
\long\def\remark#1{
        \begingroup%
        \dimen0=\columnwidth
        \advance\dimen0 by -1in%
        \setbox0=\hbox{\parbox[b]{\dimen0}{\protect\em #1}}
        \dimen1=\ht0\advance\dimen1 by 2pt%
        \dimen2=\dp0\advance\dimen2 by 2pt%
        \vskip 0.25pt%
        \hbox to \columnwidth{%
                \vrule height\dimen1 width 3pt depth\dimen2%
                \hss\copy0\hss%
                \vrule height\dimen1 width 3pt depth\dimen2%
        }%
        \endgroup%
}

\usepackage{xcolor}

\definecolor{cyanish}{rgb}{0,0.8,1.0}
\definecolor{orange}{rgb}{1.0,0.5,0.0}
\definecolor{pink}{rgb}{1.0,0.47,0.6}
\definecolor{light-gray}{gray}{0.95}
\definecolor{jiancolor}{RGB}{0,153,153}
\definecolor{mygreen}{RGB}{50,200,50}
\definecolor{pink}{rgb}{1.0,0.47,0.6}

\definecolor{commentgreen}{rgb}{0.0,0.5,0.0}

\newcommand{\boldparagraph}[1]{\vspace*{0.5ex}\noindent\textbf{#1}\hspace{1em}}

\newcommand{\ignore}[1]{}

\newcommand{\reffig}[1]{Figure~\ref{#1}}
\newcommand{\refsec}[1]{Section~\ref{#1}}
\newcommand{\reftab}[1]{Table~\ref{#1}}
\newcommand{\reflns}[2]{Lines~\hyperref[#1]{\ref*{#1}-\ref*{#2}}}

\newcommand{\x}[1]{$\times$}

\newif\ifcutforspace
\long\def\cutforspace#1{
\ifcutforspace%
        \begingroup%
        \dimen0=\columnwidth
        \advance\dimen0 by -1in%
        \setbox0=\hbox{\parbox[b]{\dimen0}{\protect{\em Cut For Space} #1}}
        \dimen1=\ht0\advance\dimen1 by 2pt%
        \dimen2=\dp0\advance\dimen2 by 2pt%
        \vskip 0.25pt%
        \hbox to \columnwidth{%
                \vrule height\dimen1 width 3pt depth\dimen2%
                \hss\copy0\hss%
                \vrule height\dimen1 width 3pt depth\dimen2%
        }%
        \endgroup%
\fi}

\newcommand{\nova}[1]{NOVA} 

\usepackage{csquotes}

\newcommand{\Mats}[1]{Mats}
\newcommand{\typedschematicblock}[1]{typed schematic block}
\newcommand{\Typedschematicblock}[1]{Typed schematic block}
\newcommand{\TypedSchematicBlock}[1]{Typed Schematic Block}
\newcommand{\interfacetype}[1]{circuit interface type}
\newcommand{\Interfacetype}[1]{Circuit Interface type}
\newcommand{\InterfaceType}[1]{Circuit Interface Type}
\newcommand{\interfacecheking}[1]{type checking}
\newcommand{\Interfacecheking}[1]{Type checking}
\newcommand{\InterfaceChecking}[1]{Type Checking}
\newcommand{\interactiveblock}[1]{visual block}
\newcommand{\Interactiveblock}[1]{Visual block}
\newcommand{\InteractiveBlock}[1]{Visual Block}

\def\paperFigure[#1,#2,#3,#4,#5]{
    \begin{figure}
        \centering
        \includegraphics[#1]{#2}
        \caption[]{#3} \label{#4}
        \Description{#5}
    \end{figure}}

\def\paperFigureWide[#1,#2,#3,#4,#5]{
    \begin{figure*} 
        \centering
        \includegraphics[#1]{#2}
        \caption[]{#3} \label{#4}
        \Description{#5}
    \end{figure*}}

\AtBeginDocument{%
  \providecommand\BibTeX{{%
    \normalfont B\kern-0.5em{\scshape i\kern-0.25em b}\kern-0.8em\TeX}}}

\setcopyright{acmcopyright}
\copyrightyear{2018}
\acmYear{2018}
\acmDOI{10.1145/1122445.1122456}

\acmConference[Woodstock '18]{Woodstock '18: ACM Symposium on Neural
  Gaze Detection}{June 03--05, 2018}{Woodstock, NY}
\acmBooktitle{Woodstock '18: ACM Symposium on Neural Gaze Detection,
  June 03--05, 2018, Woodstock, NY}
\acmPrice{15.00}
\acmISBN{978-1-4503-XXXX-X/18/06}

\begin{document}

\title{TypedSchematics: A Block-based PCB Design Tool with Real-time Detection of Common Connection Errors.}

\author{J. Garza}
\affiliation{%
  \department{Computer Science and Engineering}
  \institution{UC San Diego}
  \city{San Diego}
  \state{CA}
  \country{USA}
}
\email{jgarzagu@eng.ucsd.edu}

\author{Steven Swanson}
\affiliation{%
  \department{Computer Science and Engineering}
  \institution{UC San Diego}
  \city{San Diego}
  \state{CA}
  \country{USA}
}
\email{swanson@eng.ucsd.edu}
\date{}

\renewcommand{\shortauthors}{Garza, et al.}



\begin{teaserfigure}
  \centering
  \includegraphics[width=0.95\textwidth]{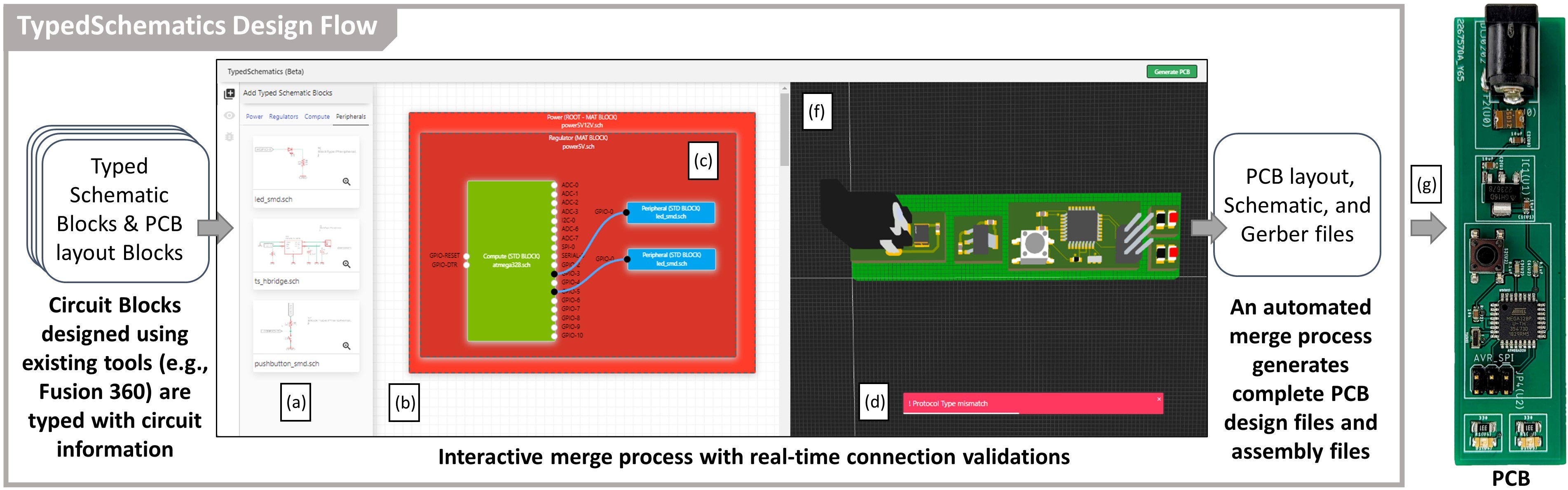}
  \caption{
    In the TypedSchematics design flow, the community creates reusable circuit blocks using existing PCB design tools. Typed schematic blocks and PCB layout blocks are imported to our interactive tool (a). Typed schematic blocks are dragged and dropped to a workspace area (b). A simplified schematic representation model reduces the number of interactions required to connect schematic blocks (c). Real-time connection error validation allows schematic blocks to be connected safely (d). Interactive 3D PCB Blocks are placed in a virtual PCB surface (f). TypedSchematics tool automatically merges both schematic and PCB blocks to create a complete PCB design (g).
  }
  \label{fig:teaserfigure}
  \Description{A linear flowchart with four elements connected with arrows pointing to the right. First multiple blocks grouped together representing typed schematic blocks and PCB blocks. Second, our TypedSchematics graphical interface. Third, a single block representing the automated process. These three blocks are steps involved in the TypedSchematics design flow. Finally, the fourth block, the manufactured PCB design.}
\end{teaserfigure}

\begin{abstract}
    Within PCB design, the reuse of circuit design blocks is a major preventing factor inhibiting beginners from reusing designs made by experts, a common practice in software but non-existent in circuit design at large. Despite efforts to improve reusability (e.g. block-based PCB design) by platforms such as SparkFun ALC and Altium Upverter, they lack merging techniques that safely guide users in connecting different circuit blocks without requiring assistance from third-party engineers. In this paper, we propose TypedSchematics, a block-based standalone PCB design tool that supports beginners create their own PCBs by providing a language syntax for typing circuit blocks with circuit data that addresses multiple challenges, from real-time detection of connection errors to automated composition and user-scalable libraries of circuit blocks. Through a user study, we demonstrate TypedSchematics improvements in design support for merging circuit blocks compared to Fusion 360. Three PCBs designed with TypedSchematics further showcase our tool capabilities, one designed by high school students demonstrates the potential of TypedSchematics to significantly lower the PCB design skill-floor.

\end{abstract}

\maketitle

\section{IntrodUction}
\label{sec:introdUction}

The de facto process of creating PCBs involves first designing an electrical schematic, a simplified version of the circuit connections, followed by a PCB layout where actual component sizes and physical connections are made. In the schematic design process, users place virtual electronic component symbols on virtual sheets and connect their signals together. In the PCB layout process, designers make the same connections but with different sized traces to electrical component footprints that come in irregular sizes. The PCB layout design is then sent to manufacturing which results in the green cards we all know. KiCad, Altium, EasyEDA, and Fusion 360 (examples of popular PCB Design tools), often provide methods for creating circuit design blocks as well as reusable circuit designs composed of both schematics and PCB layout design blocks.

While the drop in the cost of PCB manufacturing has made hardware design more accessible, opening up professional-level PCB design to students, hobbyists, and makers, the steep learning curve of PCB design tools and the lack of simpler alternatives keeps PCB design a challenge. The growing community of circuit designers often express the process of reusing and merging circuit design blocks as one of the main challenges with PCB design tools~\cite{lin2019beyond}. While some current PCB design tools provide techniques for reusing circuit design blocks, such as hierarchical sheets or device sheets, these reuse techniques are aimed at designers familiar with those blocks. Hence, when it comes to exporting and importing these reusable blocks, PCB tools leave designers with limited to no support on how to connect these blocks back together. This makes merging or composing reusable circuit design blocks challenging for beginners and experts unfamiliar with the blocks, resulting in circuit designers often resorting to create electrical schematics and PCB layouts from scratch. Even experienced circuit designers using existing circuit design blocks must carefully analyze each design they want to combine, a process that requires a lot of time and expertise and very often leads to connection errors.

Commercial block-based PCB design tools such as Sparkfun's À La Carte (ALC)~\cite{SparkfunALC} and Altium's Upverter Modular~\cite{AltiumUpverterModular} have taken steps to improve the reuse of circuit blocks for beginners. These tools provide easier to use graphical interfaces and focus mainly in improving the reusability challenge of circuit blocks. Block-based PCB design tools work by guiding designers in the selection of different pre-tested circuit blocks, such as microcontrollers, sensors, and connectors, that can be combined to create complete new designs. Forms of connection validations exist in some of these tools, such as with Altium's Upverter Modular. However, with these tools, third-party expert engineers are required to fully check for connection errors, manually merge the different PCB and schematic design blocks, or completely route the PCB design, design fees that can reach up to \$1000 per design~\cite{SparkfunALCFaq}. Furthermore, a fundamental challenge with these block-based PCB design tools is that the circuit blocks can only be scaled by the designers of the tools and not by the users, limiting the availability of circuit blocks to only a few. To the best of our knowledge, there is currently no standalone block-based PCB design tool that can perform connection validations, automatically create a fully routed PCB design, and allow the circuit block library to be scalable by users.


When it comes to merging different circuit design blocks together, standards that provide more information about the circuit block IO's and capabilities are almost non-existent in traditional PCB design tools. Comparing the current state of reusability in existing PCB design tools with software design would be equivalent to software designers manually figuring out, by reading the entire code, what data types (e.g., bool, int, char) are needed as parameters for each function code block to merge with existing code. Data type information in software is essential to automatically merge code together using compilers and also makes it possible for assistive programs detect data type errors at compile time or real-time, also known as type checking. In circuit design, the concept of data types does not exist; the closest related data types for circuit design are interface protocols that specify how to transfer data (e.g., Serial, I2C, or PCIe), and naming conventions on nets for voltages and electrical currents. However, despite providing some guidelines for circuit designers on how to make connections, these interfaces and naming conventions lack methods that can provide type checking mechanisms to validate circuit connections. Creating data types for circuits is necessary to develop both type-checking mechanisms that validate circuit connections and tools that can automate the process of merging circuit blocks, similar to compilers. 

While code-based PCB design tools have explored methods for adding circuit data types for automatically merging schematic and PCB layout design blocks and forms of type checking ~\cite{garza2019amalgam, garza2021appliancizer, lin2020polymorphic,jitpcb}, these tools are not designed for very beginners as they require additional expertise by requiring programming knowledge to recreate circuit designs using code. Code-based PCB design tools also have difficulty integrating into the design flow that predominantly consists of interactive visual tools. To the best of our knowledge, there is currently no block-based PCB design tool that integrates the benefits of adding circuit data types to its system.


In this paper, we present \emph{TypedSchematics}, a block-based standalone PCB design tool built around a syntax language that solves three design challenges with a single approach: integration of complex connection validations, automated merging of circuit blocks, and scalability of the circuit blocks library by users.


With TypedSchematics, expert users begin annotating schematic blocks with circuit data type information and create a PCB block for the schematic design using current PCB design tools. These circuit blocks, composed of schematic and PCB blocks, are then imported to our tool and reused by beginners. A standardized syntax language for declaring circuit data types is followed during the annotation process. The annotated schematic and PCB design blocks are then imported into our interactive design tool. TypedSchematics guides beginners and experts alike during the connection process of wires between circuit blocks by showing errors and warning in real-time. An automated merging process generates Gerber files, manufacturing-ready files for the fabrication of the PCBs.

TypedSchematics is inspired by Blockly~\cite{Blockly}, a block-based software design tool that provides an alternative to more complex software programming languages. Like Blockly, our tool is intended as a beginner-friendly alternative to PCB design tools. Our tool is not intended to compete with the more sophisticated and robust traditional PCB design tools designed for experts.

In contrast to existing commercial solutions, such as Sparkfun ALC and Altium Upverter, which require external engineers to merge the circuit blocks or manual routing, our tool's merge process is automated and generates schematics, routed PCB designs, and Gerber files, without need for human intervention. Unlike the tools previously mentioned, our tool allows users to import more circuit blocks to our library. Compared to code-based PCB design tools, TypedSchematics leverages the already-established visual design flow for better adoption between beginners.

Block-based PCB design tools are characterized by providing visual and interactive design flows using blocks; In this area, TypedSchematics also presents its own unique approach to visually connecting circuit blocks that we call \textit{"Mats"}. In TypedSchematics, Mats eliminates the need for making repetitive connections (e.g., VDD and GND), reducing interaction costs. Additionally, Mats integrates a flow-based programming paradigm that allows connection validations to be performed in real-time. 

The ultimate goal of TypedSchematics is to significantly reduce the PCB design skill-floor by providing an alternative PCB design workflow that supports beginners in design without requiring supervision or assistance from expert designers. To reduce the entry level of PCB design, TypedSchematics addresses various challenges with a unique solution that is the introduction of a syntax language that allows creating a standalone and alternative PCB design workflow, although we will do our best to detail each part of the tool including the PCB layout generation and routing, this paper focuses on the main contributions which are:

\begin{itemize}
    \item A syntax language for reusing circuit blocks that solves three design challenges: supporting beginners detect common connection errors when merging circuit blocks, enabling automated composition of circuits blocks, and allowing scalability of the circuit blocks library by users.
    \item Mats, a schematic representation model for circuit blocks with a flow-based programming paradigm that allows connection validations to be performed in real-time.
\end{itemize}

Block-based PCB design tools require third-party engineers to merge the different circuit blocks together or complete the PCB design. These engineers use traditional PCB design tools (e.g., KiCad, Fusion 360) in the background for the merge process. Our tool currently outperforms commercially available block-based PCB design tools in the sense that our tool does not require a human-in-the-loop for creating completely routed PCB designs. Therefore, to test how TypedSchematics offers a better support for merging circuit blocks, TypedSchematics compares itself directly to Fusion 360 through a user study. In this study, we also compared the simplified schematic representation model "Mats" against the schematic model for circuit blocks used in Fusion 360. Three PCB designs made with TypedSchematics further demonstrate the capabilities of our tool; One of those designs was developed by high school students without assistance.


The rest of this paper is organized as follows. \refsec{sec:related} describes previous work and techniques for reusing circuit designs and discusses existing block-based PCB design tools. \refsec{sec:challenges} explains some design challenges with existing schematics and PCB design reuse techniques.
\refsec{sec:overview} introduces TypedSchematics. 
\refsec{sec:examples} demonstrates our tool capabilities by describing three devices made with TypedSchematics.
\refsec{sec:userstudy}, \refsec{sec:results}, and \refsec{sec:discussion} demonstrates how TypedSchematics addresses the design challenges. \refsec{sec:limitations} presents applications, limitations and future work and \refsec{sec:conclude} concludes.



\section{Related Work}
\label{sec:related}

Our work focuses on a block-based PCB design tool that improves the reuse of circuit blocks, specifically the merging process, by adding circuit data type information to schematic blocks, enabling detection of common connection errors and automated composition of both schematic and PCB blocks. We also introduce "Mats", a circuit representation model for circuit blocks that reduces design interaction costs and allows for detection of connection errors to be real-time. In this section we review previous work related to our research.

\subsection{HCI and Circuit Design}

Recently, there is a growing interest in the HCI community in tools that help and assist circuit designers in the different stages of circuit development, from prototyping to PCB design and testing. In the prototyping stage, previous research has work on projects that assist designers in the initial circuit prototyping exploration that use breadboards~\cite{kim2020schemaboard, drew2016toastboard}. Other works have focused on improving the debugging and visualization of analog and digital signals of circuits ~\cite{strasnick2021coupling, wu2017currentviz}. While other works~\cite{hodges2020democratizing, strasnick2019pinpoint} focus on the final stages of circuit design, such as supporting debugging and mass production of PCBs. Our work, on the other hand, focuses on the schematic and PCB design stages, stages that usually follow after the initial prototyping of circuits using breadboards or development boards.                          

\subsection{Traditional PCB design Tools and Design Reuse}
\label{sec:designreuse}

Traditional PCB design tools like Fusion 360~\cite{EAGLE}, KiCad~\cite{KiCad}, Altium~\cite{Altium}, and EasyEDA~\cite{EasyEDA} primarily consist of a schematic capture software and a PCB layout software. These PCB design tools have different methods for reusing schematic and PCB design blocks. Below we explain these methods and their limitations. 
\\
\boldparagraph{Hierarchical Sheets} also known as Multi-Sheet Hierarchical Design in Altium~\cite{AltiumMultiSheet} allows designers to abstract the complexity of schematics into reusable blocks. In this method, the designer creates a schematic subsheet and selects signals in the design to export. On another sheet, designers can draw one or more reusable blocks that reference the created subsheet and exported signals. The exported signals serve as inputs and outputs for the reusable block to be connected with other elements within the schematic sheet.

Altium also brings the concept of Device Sheets~\cite{AltiumDeviceSheet}, hierarchical sheets with the ability to be reused between projects, compared to Altium's Multi-sheets. Device sheets can also be versioned, called Managed Device Sheets.

Fusion 360, KiCad, EasyEDA and Altium implement some type of hierarchical sheets; however, this reuse method is limited to schematics.
\\
\boldparagraph{Design Snippets}~\cite{AltiumDeviceSheet} in Altium and similarly Modular Design Blocks~\cite{ModularDesignBlocks} in Fusion 360, its an alternative reusability method that allows both schematics and PCB layouts to be copied and pasted into other designs. 
\\
While the above methods allow for the reuse of schematic and PCB blocks, when it comes to connecting different circuit blocks together or merging them, these reuse methods face the same fundamental problem, which is that these methods do not provide clear information on how to connect the different blocks together. For example, an IO wire of a circuit block may be named by a designer as "Power", although the name implies that it could be a power signal there is no certainty that it is a power signal, let alone what voltage it supports. Without more information about these nets, designers are often clueless about how to connect these blocks together. For small teams, the above reuse methods work perfectly as they can share information on how to reuse each block; However, when it comes to reuse by the circuit design community at large of both experts and beginners, the problem becomes very apparent. 

Circuit data types or other information that guides designers in connecting circuit blocks would facilitate circuit reuse and currently do not exist in traditional PCB design tools. The lack of support when making connections between circuit blocks may be one of the reasons why circuit design reuse applying reuse methods in the hardware community at large is not as common as it is in the software community. Often, circuit design reuse among the broader circuit community involves looking at the complete design of others and copying the entire design or parts of it component by component.

Other tools such as Fritzing~\cite{knorig2009fritzing} allow visually connecting different circuits blocks; However, when it comes to creating the schematic and PCB layout of circuits, designers are left only with standard PCB design capabilities, which even lack basic reuse methods such as hierarchical sheets.

\subsection{Block-based PCB design Tools}

Block-based PCB design tools are tools that provide a design alternative compared to traditional PCB design tools that focus on easier-to-use graphical user interfaces built around reusable circuit blocks. Altium Upverter formerly Gpetto~\cite{AltiumUpverterModular} and SparkFun À La Carte (ALC)~\cite{SparkfunALC} are block-based PCB design tools.

Each tool provides different graphical user interface experiences for merging circuit blocks. Sparkfun ALC provides users with a shopping cart style graphical interface where users can select the different circuit blocks they want to combine. However, with SparkFun ALC users cannot manipulate the wire connections and the placement of PCB blocks. In contrast, Altium Upverter provides a more sophisticated user interface that allows designers to choose how to connect wires using drag-and-drop menus and select the position of PCB blocks along a PCB surface. 

For connection validations, SparkFun leaves connection validations to external engineers, while Altium Upverter provides some connection validation mechanism that allows connections between hardware protocol types (e.g., I2C to I2C). However, Altium  Upverterlacks more complex validation mechanism like checking for I2C addresses conflicts or master-slave constraints. Also, Altium hardware protocol types and validations cannot be dynamically extended by users.

The process of merging circuit blocks for both tools is also different. In Sparkfun ALC, both the schematic and PCB blocks merge processes is left to third-party engineers who use traditional PCB design tools for this process. Altium Upverter provides the ability to generate complete schematic files, but the PCB design is left up to the user or third-party engineers. In other words, these tools cannot work standalone.

Neither Sparkfun ALC or Altium Upverter have standards or methods to add new community-designed circuit blocks to the tools. The circuit blocks in both design tools are fixed and are only extensible by the developers of the tools.

\subsection{Code-based PCB design tools}

Code-based PCB design tools such as Polymorphic Block~\cite{lin2020polymorphic} and JITPCB~\cite{jitpcb}, enable reuse of circuit design blocks with the addition of hardware description languages (HDLs) for board-level circuits. In other words, these tools use code to represent the connections between electrical components or blocks of components instead of using visual circuit diagrams. An advantage of using HDLs to build circuits is the ability to easily run constraint validation solvers until the correct circuits are found that fit the system requirements. While code-based PCB design tools have many benefits in terms of automation, it requires learning a completely different set of tools and workflows (e.g., code syntax for circuits, compilation process), resulting in designers having a steep learning curve compared to the traditional use of visual design workflows.

Other work has focused on increasing the abstraction of circuit design by making use of modular design and reusing common circuit designs. Trigger-Action-Circuits~\cite{trigger-action} reuses popular circuit implementations and abstracts the design to a behavioral level and makes possible the automated generation of circuit prototypes from an if-then programming approach. Appliancizer~\cite{garza2021appliancizer}, on the other hand, abstracts the circuits to the HTML GUI level, allowing for schematics and PCB layout automated generation from web interfaces. While these tools allow the creation of circuit designs almost automatically, the high abstraction eliminates the ability to configure low aspects of the design, for example, pin and wire connection selections between circuit blocks are impossible in these tools. 

Similarly, Trigger-Action-Circutis and Appliancizer require developers to have knowledge of other areas to create and add additional blocks. For example, Trigger-Action-Circuits block developers require XML knowledge, while Appliancizer block developers require HTML/CSS knowledge for adding additional circuit design blocks.

Unlike these tools, our tool works at a lower level allowing designers to manually define the connections between circuit blocks. Our technique for encapsulating circuit data information is integrated into existing PCB design tools, simplifying development by not requiring circuit block developers to learn new programming languages, making the circuit block's library more scalable and easier to adopt. 

\subsection{Visual Schematic Representation Models}

The visual representation model of schematics consists of connections of circuit symbols (e.g., resistors, capacitors, integrated circuits) and uses a graph structure algorithm, using nodes to represent symbols and edges for possible connections. This visual model has not changed much since the standardization of circuit symbols~\cite{SchematicSymbols} and the introduction of the first schematic capture software~\cite{FirstPCBSoftware}.

However, the search for alternative circuit design approaches for new use cases has led to rethinking the representation of the schematic model for specific applications that bring added benefits to their users. Aesthetic Electronics~\cite{lo2016aesthetic}, for example, combines the schematic model with graphic design tools bringing improvement to the fabrication of circuits embedded in fabrics or ornaments. Appliancizer~\cite{garza2021appliancizer} combines web design (HTML, CSS, JS) with circuit abstractions for the rapid creation of gadgets that feature screens. Polymorphic blocks ~\cite{lin2020polymorphic} uses code as the basis of representation of the schematic model which allows the automation of circuit designs. Our tool adds enhancements to the visual schematic model to facilitate the development specifically for circuit blocks.
\section{Design Challenges Using Reusable Circuit Blocks}
\label{sec:challenges}

Schematic and PCB layout design reusing circuit blocks presents unique challenges compared to designs made from scratch. The current merging process with schematic blocks lacks standards to provide information of the circuit blocks to merge, leaving designers in the dark when it comes to connecting blocks made by other designers. Below, we describe in more detail some of the challenges that arise when merging schematic and PCB blocks.

\subsection{Hardware Protocol Connection}

Hardware protocols (e.g., I2C, SPI, GPIO) have become a popular way for transmitting information between ICs (Integrated Circuits). These hardware protocols range from a single wire (e.g., GPIO) to many wires (e.g., SPI), and each protocol has different special ways of connecting these wires. A single mistake connecting one of these wires can produce a circuit error. For example, the I2C hardware protocol which consists of two wires named Serial Clock (SCL) and Serial Data (SDA), cannot mix its wires (e.g., SCL with SDL). Knowing how to connect these protocols creates the first challenge when combining schematic blocks, requiring designers to know these protocols properly in order to connect them correctly.

Furthermore, these protocols also integrate different capabilities making them have connection restrictions in special circumstances. One example is the hardware protocol SPI which uses three wires to transmit data, one clock wire (SCK) and two data wires (MISO, MOSI). This protocol requires one IC to be configured as Master and another IC as Slave, making other configuration connections impossible otherwise. Similarly, the I2C protocol requires each IC to have a unique 7-bit address (e.g., '0\x{}1F'), making duplicate I2C addresses connections impossible. If any of these extra capabilities are not met, the circuit fails. This information is rarely included in schematic designs.

\subsection{Supply Voltages}

All schematic blocks require at least a voltage and ground to power the block's internal circuitry. The required voltage may vary depending on the schematic block design, with some requiring a fixed voltage (e.g., 3.3V, 5V) and others allowing a range of voltages (e.g., 3.3V-5V), depending on the block's internal circuitry capabilities. Existing schematic design tools also lack of methods to include voltage information in schematic blocks, making finding the required block's voltage another design challenge. Since all schematic blocks require power and ground signals, connecting these wires becomes a repetitive task, which often resulting wire clutter in the design.

\subsection{Unknown Signals and Optional Connections}

Designers generally try to annotate schematic block's input and output signals with meaningful names so that it is understood if these signals are part of a protocol (e.g., SDA, SCL) or not. Since there are no standards for annotating signals, the block signals and their capabilities are generally unknown. An unknown signal example might be a signal named "Reset", which although the name is meaningful, does not indicate whether it is part of any hardware protocol or its connection capabilities.

\subsection{PCB blocks design shape and traces}

When it comes to exporting and importing PCB layout designs, there is no standard that defines the block design shape or whether the user must leave certain traces routed or not. This usually makes it impossible for the designer to merge PCB layouts into new designs easily. TypedSchematics presents unique views for exporting and merging PCB layouts which address these challenges.
\section{TypedSchematics}
\label{sec:overview}

To better support designers in merging circuit blocks, we created TypedSchematics~\footnote{
    
TypedSchematics is available online at \url{https://typedschematics.com/} 

}. TypedSchematics is a block-based PCB design tool that introduces a circuit blocks reusability method that features effortless integration with other circuit blocks by providing detection of common connection errors. An automated composition of circuit blocks allows the generation of full schematics, PCB layouts and PCB manufacturing files, removing third-party engineers from the design loop. In contrast with exiting block-based PCB design tools, TypedScheamtics circuit blocks library is user-scalable. TypedSchematics' detection of common connection errors, automated circuit block composition, and user-scalable circuit block library are made possible with a language syntax that allows the declaration of circuit data types that are annotated in schematic blocks using existing PCB design tools.

TypedSchematics also improves the schematic representation model for designing circuits using circuit blocks with the introduction of Mats, interactive design blocks with grouping capabilities, which allows the automatic connection of power signals, reducing interaction costs. Mats also futures a flow-based programming paradigm that enables the detection of common connections errors be real-time. TypedScheamtics implementation is web-based. Below is an overview of the design process with our tool, followed by the language syntax and system design. 



\subsection{Design Process Overview}

\reffig{fig:teaserfigure} gives an overview of the design flow with our tool. With TypedSchematics, expert designers start by adding circuit data type information to schematic blocks created with existing PCB design tools (e.g., Fusion 360), creating with this \emph{\typedschematicblock{}s}. The type information follows a minimal syntax of symbols and naming conventions for declaring circuit interfaces that are then written in signal net names or as text in schematic blocks. \Typedschematicblock{}s, and the corresponding PCB layout design for each schematic block, are added to our interactive tool and displayed in a sidebar for both experts and novice designers to use (a). 

\Typedschematicblock{}s are classified into four groups based on their capabilities: power, regulators, and computer peripherals. This separation helps designers better understand the core capabilities of each block. Designers then add \typedschematicblock{}s to a workspace area (b) where each \typedschematicblock{} is rendered as a visual interactive block (c), each with different interactive capabilities depending on their classification. The rendered blocks are connected using our schematic representation model that we call \emph{Mats}. Mats allows grouping compute and peripheral blocks inside power and regulator blocks, avoiding having to connect ground and voltage signals for all schematic blocks, making design with schematic blocks faster and clearer.

For each schematic block connection, our tool runs an \interfacecheking{} algorithm in real-time, which validates voltages and interfaces connections, among others. In case of a connection error, a message is displayed to designers indicating the cause of the error (d). A 3D editor allows designers to place 3D PCB blocks on a virtual PCB surface, giving an idea of what the final PCB design would look like (f). After designers finish connecting and placing the desired schematic blocks and PCB layout blocks, with a simple click, our tool generates complete schematic and PCB layout files, as well as Gerber files, allowing designers to make adjustments to generated schematic of PCB layout design files or send the produced PCB design directly to manufacturing using the Gerber files. A PCB design after manufacture is shown in (g).

\subsection{\TypedSchematicBlock{}s}

\Typedschematicblock{}s are electronic schematics annotated with type information using our own language. Similar to programming languages, this language allows circuit designers to declare types within schematic blocks using existing PCB design tools (e.g., Fusion 360). While programming languages use data types to classify data with different attributes and data sizes (e.g., bool, int, char), our tool proposes the use of hardware interface protocols (e.g., GPIO, I2C, SPI) and power signals (e.g., 3.3V, 5V) for type declarations. We call type declarations for schematic blocks as \textit{\interfacetype{}s}. \Interfacetype{}s declarations are made directly on signal nets of schematic diagrams. The following sections briefly explain the annotation process to create \typedschematicblock{}s.




\subsection{Language syntax}
\label{sec:typesyntax}

The language syntax for \interfacetype{}s declarations is divided in three categories. The first protocol \interfacetype{}s which are used as connection edges for the blocks and define circuit protocol connections (e.g., SPI, I2C, GPIO). The second power \interfacetype{}s which describe the power capabilities of the blocks (e.g., 3.3V, 5V). Finally, a global attributes syntax to declare further block information, as well as added capabilities of \interfacetype{}s. 

Our language syntax was based on common patterns followed by designers to annotate signal names, making it easier for designers to understand the syntax. Since space for declaring types is short, our syntax is highly symbolic.  \reftab{tab:SyntaxStyleTable} shows the full syntax style for declaring \interfacetype{}s, below we briefly explain each syntax category.

\begin{table}
    \caption{\TypedSchematicBlock{}s Language Syntax}
    \label{tab:SyntaxStyleTable}
    \begin{tabular}{ccl}
        \toprule
        Types             & Syntax                       \\
        \midrule
        Protocol          & \#{Protocol.Signal\_Voltage} \\
        Power             & @{(VIN or VOUT)\_Voltage}    \\
        Power             & GND                          \\
        Global Attributes & \#{Global Attributes}        \\
        \bottomrule
        Optional Symbols  & Explanation                  \\
        \midrule
        -                 & Alternate name               \\
        !                 & Protocol not required        \\
        \midrule
    \end{tabular}
\end{table}

\subsubsection{Protocol \InterfaceType{}s}

Protocol \interfacetype{}s are declared with the '\#' ASCII prefix followed by the protocol name (e.g., SPI, I2C, GPIO). Protocol \interfacetype{}s can consist of multiple signals. The Inter-Integrated Circuit (I2C) protocol is an example, this protocol is made up of two signals usually named Serial Clock Line (SCL) and Serial Data Line (SDA). The symbol '.'  can be used to group signals of the same protocol (e.g., \#I2C.SDA, \#I2C.SCL). Our tool displays protocol signals as one wire, avoiding the need for users to connect each protocol signal one by one.

It is common for microcontrollers or integrated circuits to have more than one same protocol. The ASCII '-' symbol gives the interface protocol an alternate name distinguishable from the others, allowing the enumeration or description of functionality (e.g., \#GPIO-0, \#GPIO0-1, \#GPIO-RESET).

Finally, it is also common in schematic block designs to leave some connections as optional. The symbol '!' allows to indicate that the connection of a protocol \interfacetype{} is optional (e.g., \#GPIO-0!).

\subsubsection{Power \InterfaceType{}s}

Power \interfacetype{}s are declared with the '@' ASCII prefix followed by two reserved words 'VIN' or 'VOUT' which indicates our tool if the schematic block receives power, outputs power, or both. If the \typedschematicblock{} outputs power, it is classified as a power block. The classification of the different blocks is covered in detail in \refsec{sec:blocksClassification}.

Following the previous syntax, the ASCII '\_' symbol is used to indicate the voltage restrictions of the power signals, which allows designers to specify fixed voltages or a range of allowed voltages (e.g., \@VOUT\_3V, \@VIN\_5V-9V). By knowing the input and output voltages constraints, voltage validation is achieved.

Finally, The word 'GND', a common abbreviation for ''Ground'', is used as a reserved keyword. Our tool grounds all schematic blocks via signals declared as 'GND', when generating the full schematic.

\subsubsection{Global Attributes}

Our language syntax also allows designer to include global attributes for schematic blocks such as the block classification (e.g., Power, Peripheral) as well as additional information for the interface protocols such as the address of the I2C protocol \interfacetype{} or whether the SPI protocol is master or slave, allowing for more refined validations. These global attributes are defined within the schematic as text comments using the syntax '\#{ <global attributes> }'. 




\subsection{Block-based schematic design}
\label{sec:blocksClassification}

\Typedschematicblock{}s are imported directly into our block-based PCB design tool, TypedSchematics. A validation algorithm verifies the consistency of the files and the declared circuit types, if accepted, the imported blocks are rendered as interactive \Interactiveblock{}s when added to the workspace area.

The rendering of each \interactiveblock{} varies according to the \typedschematicblock{} classification, which can be: Power, Regulators, Compute Modules, or Peripherals. Each classification is explained below.
\\
\\ 
\boldparagraph{Power} blocks are capable of supplying power to other blocks and are the root of the tree structure that contains the entire schematic design, described in more detail in \refsec{sec:interactiveCapabilites}. \reffig{ fig:powerroot} left, shows an example of a visual power block, on the right its corresponding \typedschematicblock{} consisting of a power connector, a resettable fuse for short circuit protection and a voltage stabilizing capacitor.
\paperFigure[%
    {width=\columnwidth},
    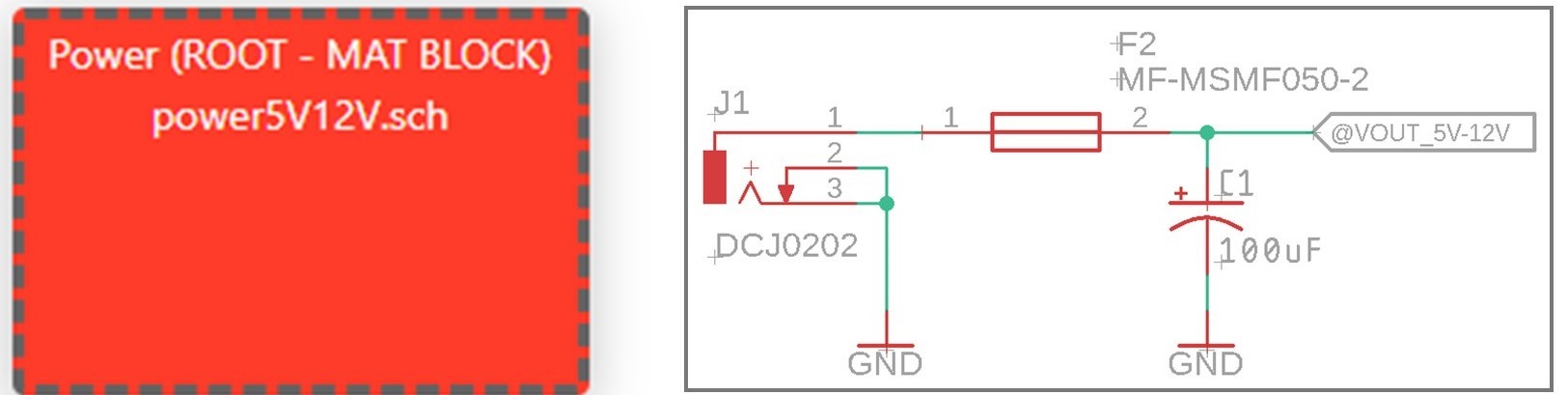,
    {A power supply connector \typedschematicblock{} (right) and its \interactiveblock{}, a power block (left).},
    fig:powerroot,
    Two blocks, on the left the visually rendered power block, a red square with the text Power, on the right the power block schematic.]
\\
\\
\boldparagraph{Regulator} blocks step the voltage up or down to supply blocks with their required voltage.\reffig{ fig:powerregulator} left, shows an example of an interactive visual regulator block, on the right its \typedschematicblock{} consisting of a 5-volt AMS1117 IC~\cite{AMS1117} power regulator. The input voltage range was taken directly from the power IC datasheet.
\paperFigure[%
    {width=\columnwidth},
    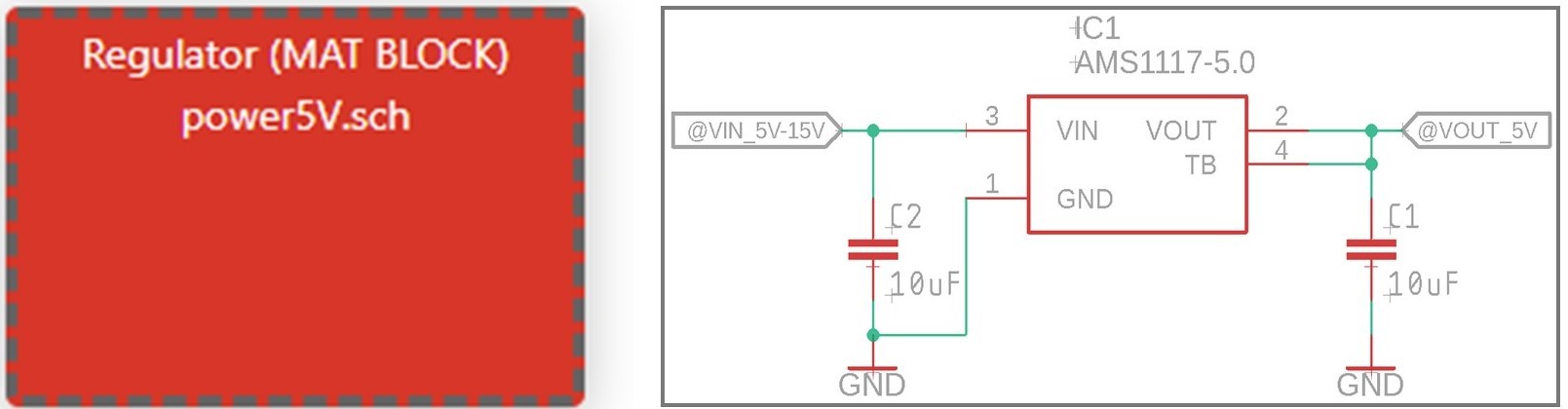,
    { A 5V regulator \typedschematicblock{} (right) and its \interactiveblock{}, a regulator block (left).},
    fig:powerregulator,
    Two blocks, on the left the visually rendered regulator block, a red square with the text Regulator, on the right the regulator schematic.
]
\\
\\
\boldparagraph{Compute module} blocks are intended for microcontrollers or processors capable of communicating with peripherals. These blocks require an input voltage and have multiple protocol \interfacetype{}s to address a variety of sensors and actuators. \reffig{ fig:computemodule} left, shows an example of a rendered compute module block, on the right its corresponding schematic consisting of an 8-bit Atmega328~\cite{Atmega328} microcontroller and accompanying electrical components: an 8~MHz crystal, programming pin headers and a reset switch, commonly used in designs.
\paperFigure[%
    {width=\columnwidth},
    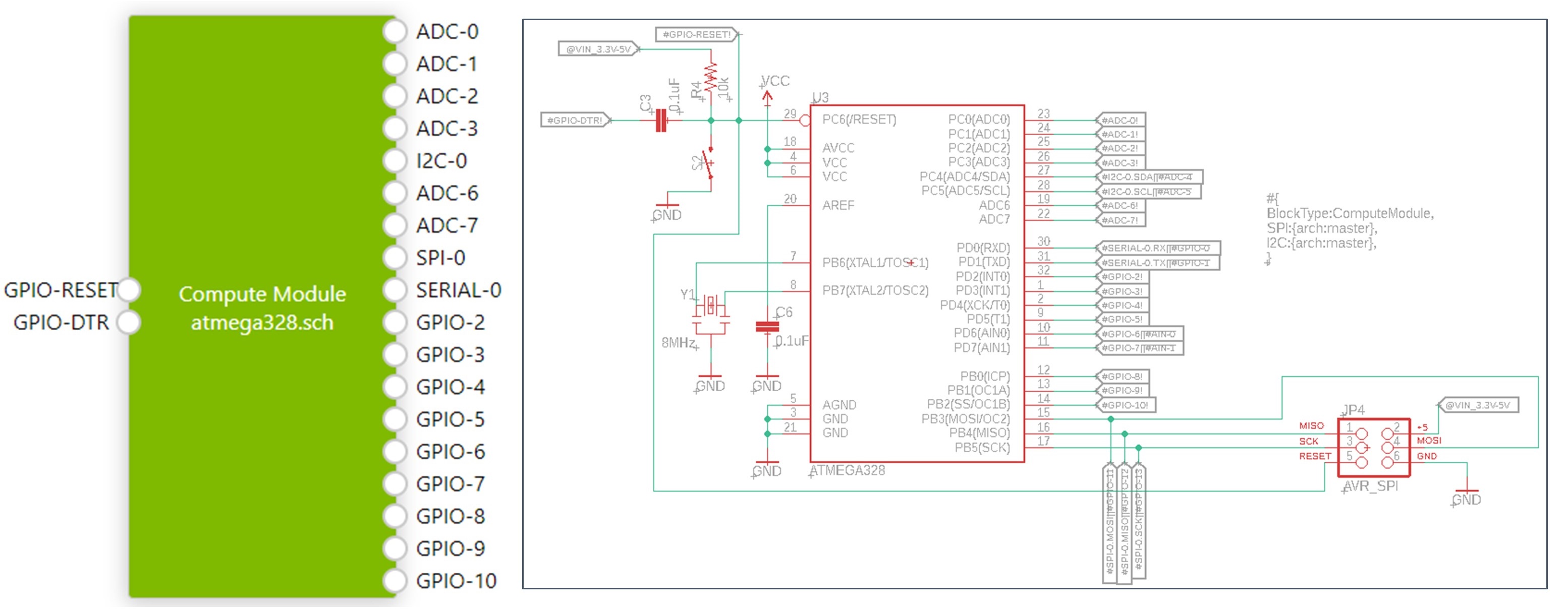,
    {An Atmega328 microcontroller \typedschematicblock{} (right) and its \interactiveblock{}, a compute module block (left).},
    fig:computemodule,
    Two blocks, on the left the visually rendered compute block, a green rectangle with the text Compute and I/O connections on the right and left edges, on the right the peripheral schematic.]
\\
\\
\boldparagraph{Peripheral} blocks are intended for sensors or actuators. These blocks also require an input voltage and have fewer protocol \interfacetype{}s compared to compute module blocks. \reffig{ fig:peripheral} left, shows an example of a rendered peripheral block, on the right its corresponding schematic consisting of a temperature sensor IC MCP9808~\cite{MCP9808} with a fixed I2C address '0\x{}18' declared as an I2C extra property using the global attributes comment block.
\paperFigure[%
    {width=\columnwidth},
    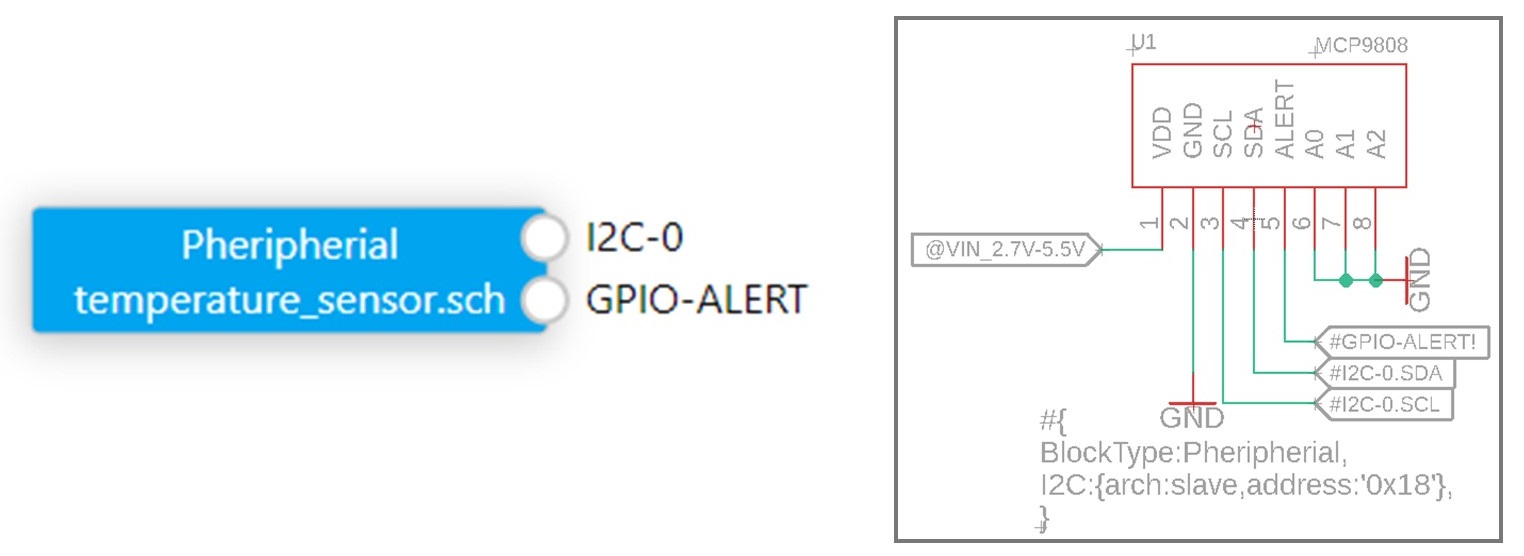,
    {A temperature sensor \typedschematicblock{} (right) and its \interactiveblock{}, a peripheral block (left).},
    fig:peripheral,
    Two blocks, on the left the visually rendered peripheral block, a blue rectangle with the text Peripheral with I/O connections on the right edge, on the right the peripheral schematic.
]

\subsection{Mats Schematic Model Representation}
\label{sec:interactiveCapabilites}

We call our schematic model representation structure for connecting circuit blocks as \emph{Mats}. Compared to the schematic model representation of conventional schematic design tools, that use a general graph structure to interactively connect all electrical symbols, Mats divides the structure into two, a tree graph structure and a graph structure.  

Each TypedSchematics visual block (power, regulator, compute modules, and peripheral) connects as a tree graph or graph. The compute module and peripheral blocks are connected as graph nodes, allowing for multiple edge connections. We call these blocks \emph{general blocks}. The power and regulator blocks are connected as tree graph nodes, we call these blocks \emph{mat blocks}. 
s
Mat blocks and general blocks have different interaction capabilities. General blocks can form connections using edges, while Mat blocks allow any block to sit on them, hence the name Mat. 

Internally, when any type of block is dragged and dropped onto a mat block, the output voltage of the mat block is connected to the input voltage of the dropped block. The main power source for the circuit is the root mat block node and distributes power to the children, the regulator blocks. 

When any type of blocks is dragged and dropped on top of a mat block, the output voltage of the mat block is connected to the input voltage of the dropped block. The power and regulator blocks form a tree with the global power supply block as its root. As an example, \reffig{ fig:treegraph} (b) shows a power block containing two regulator blocks, one 5 volts and one 3.3 volts, (a) shows the tree graph structure representation for these three mat blocks. The graph structure is exemplified in (b), where (b) shows a compute module block connected to three peripheral blocks; two LEDs, and a temperature sensor, while (d) shows the visual representation of the internal graph structure.

\paperFigure[%
    {width=\columnwidth},
    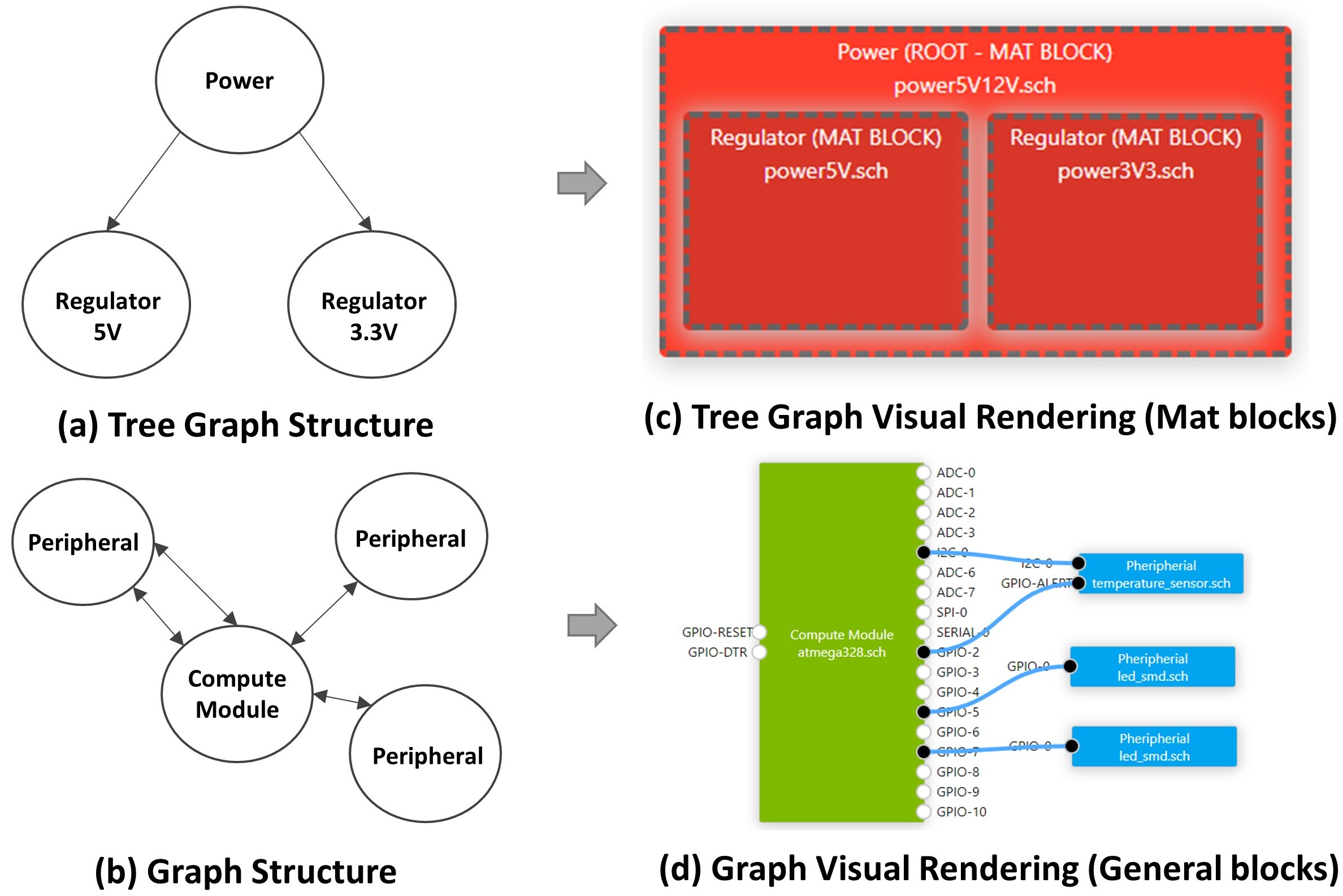,
    {A three-graph structure (a) and its visual representation as mat blocks (c) compared to the general graph structure (b) and its visual representation as general blocks (d).},
    fig:treegraph,
    Four figures, on the top right a large red power block rectangle and two smaller red regulator block rectangles inside. On the top left, three circles, one above with the text power and two below with the text Regulator connected by arrows pointing down. On the bottom right a green compute module block connected to three blue peripheral blocks. On the bottom left, four circles connected as a graph, the one in the center with the text compute module and the three around with the text peripheral.]

Mats allows for the elimination of manual power connections required by all \typedschematicblock{}s that would otherwise have to be made if represented by a single graph structure. This simplifies the visual diagram of \typedschematicblock{}s by reducing clutter, and reducing the cost of interactions by freeing the designers from having to specify the input voltages for each block.

\subsection{Mats Real-time \InterfaceChecking{}}
\label{sec:constraints}

When making connections between schematic blocks, either by connecting wires between general blocks or by dragging and dropping blocks onto mat blocks, the code within Mats checks if the connections are valid through \interfacecheking{}. \Interfacecheking{} runs in real time. To perform \interfacecheking{} in real time, a flow-based programming paradigm is seamlessly integrated into the tree and general graph structures described in \refsec{sec:interactiveCapabilites}. Consequently, when designers make connections, \interfacetype{} information is acquired using a message passing algorithm and validations are performed at once.

Four types of \interfacecheking{}s described below are supported with TypedScheamtics, an error is displayed to the designer when the constraints are not met. 

\subsubsection{Protocol Interface Matching} A protocol matching \interfacecheking{} is triggered when connecting wires between general blocks. This validation verifies that protocol interfaces are of the same type, an error is displayed in case of protocol types mismatch.
\subsubsection{Voltage Interface Matching} Voltage matching \interfacecheking{} is triggered when blocks are dragged and dropped onto mat blocks. This validation verifies that the voltage \interfacetype{} of the dropped block element and that of the mat blocks fits together. For example, if the designer wants to add a compute module block that requires 5V to a mat block with an output voltage of 3V, the action is not allowed. 
\subsubsection{Required Interface}  With the use of the '!' symbol, described in \refsec{sec:typesyntax}, designers can indicate whether protocol \interfacetype{}s are required or not. This \interfacecheking{} verifies, when attempting to generate the full schematic, that no required interface connections are missing.
\subsubsection{Additional Interface Checkings}  With the use of global attributes, described in \refsec{sec:typesyntax}, additional capabilities can be added to interface types. Interface checking for additional interface type capabilities can be extended by designers. For example, the I2C protocol interface type includes address validations, if there are duplicate addresses (i.e., two sensors with the same I2C address) the connection is invalid. Another example is the SPI protocol interface type, which includes validations that require a master-slave connection, rendering master-master or slave-slave connections invalid.

\subsubsection{Design Implementation} Mats GUI and flow-based programming paradigm integration for real-time connection validations were both designed from scratch using TypeScript, JavaScript, and HTML. Mats works as part of TypedScheamtics frontend.

\subsection{Electrical rule checks and \InterfaceChecking{}}

\InterfaceChecking{} and Electrical Rule Checks (ERC) share similarities but perform different design checks. ERCs lack notion about protocol interfaces (e.g. I2C), master-slave validations, optional connections, among others, therefore ERCs cannot be used to verify connections between circuit blocks. \InterfaceChecking{}, on the other hand, can validate connections between circuit blocks since it has notion of protocol interfaces and other interfaces from the type declarations made using the language syntax described in \refsec{sec:typesyntax}. While ERC understands voltages, \interfacecheking{} allows a range of voltages in the form of a variable whose voltage is only known after the connection between circuit blocks is made.

\InterfaceChecking{} and ERC occur at different design stages. \InterfaceChecking{} occurs at the time of making connections between circuit blocks. After the schematic and PCB blocks are automatically merged, a process described later in \refsec{sec:automatic}, an ERC is run to further validate the integrity of the circuit design.

When \typedschematicblock{}s are imported to our tool, our tool also runs an ERC to validate the integrity of the schematic block being imported together with a syntax validation of the type declarations. ERCs are not always reliable; therefore, expert designers who load \typedschematicblock{}s must be confident that the design is error-free.

The same applies to PCB blocks. For PCB blocks, a design rules check (DRC) is run when the file is imported along with its corresponding schematic block and after the PCB blocks are automatically merged.

\subsection{Block-based PCB layout editor}

TypedSchematics also provides an interactive 3D editing tool for merging PCB blocks. These PCB blocks are created and uploaded along with their corresponding schematic blocks by expert designers. The design of each PCB block is made to fit into the smallest possible rectangular space and following design recomendations. TypedScheamtics currently only supports typed schematic and PCB blocks designed with Fusion 360. These Fusion 360 schematic and board files are imported directly into our TypedScheamtics tool for use by beginning designers, as mentioned in previous sections.

The PCB blocks, which correspond to each schematic block, are displayed in the editor as 3D virtual PCB blocks that can be placed along an empty 3D virtual PCB surface. Minimal error checks are applied during the PCB layout design process, such as checking if PCB blocks are outside the boundaries of the PCB surface. By using the 3D PCB layout editor, designers can quickly get an idea of the final PCB design product after fabrication.

\subsubsection{Design Implementation} The PCB blocks 3D editor was designed using Three.js~\cite{ThreeJS} and works as part of TypedScheamtics frontend. PCB blocks are imported to our tool as both step files and board files created using Fusion 360. For rendering the step files, the library Buerli~\cite{Buerli} was used. 

\subsection{Automated Composition of Circuit Blocks}
\label{sec:automatic}

After beginner designers place each PCB block at the desired location along the virtual PCB surface and connect each schematic block using the flow-based schematic design for circuit blocks, Mats, if there are no design errors, TypedSchematics automatically merges the schematic and PCB blocks into a single design. For merging the schematic blocks, the tree and general graph structures are converted into a single graph and the signals of each \interfacetype{} are connected according to the design. For merging the PCB layout blocks, the different PCB layout designs are copied into a single PCB layout design, and the missing trace connections are auto routed. Along with individual schematics and PCB layout files, TypedSchematics also automatically generates Gerber files for PCB manufacturing.

The algorithms for merging schematics and PCB blocks is extensive but straight forward. In short, the algorithm copies each selected schematic and PCB block into a single file and uses the resulting Mats wire connections to connect each block. One of the complexities of the algorithm is resolving duplicate PCB schematics and blocks, since their reference designators and signal names are the same. Our tool addresses this problem by adding prefixes to each reference designator and signal name. To highlight each block in the final design, we include the edge of each block in the silkscreen which appears as white border squares on the final PCB.

\paperFigure[%
    {width=\columnwidth},
    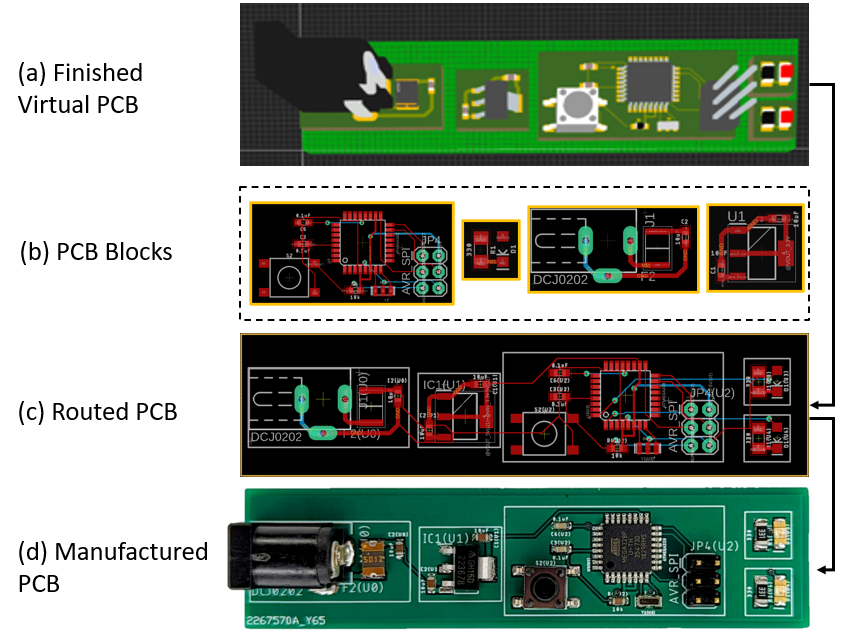,
    { Once a virtual PCB is completed (a), user-selected PCB blocks (b) are automatically merged into a single file and missing connections are routed (c), (d) shows the final fabricated PCB.},
    fig:pcbblocks,
    Four figures arranged from top to bottom. At the top, a virtual PCB example circuit, then some PCB blocks, followed by a routed PCB layout of the example circuit containing the PCB blocks. At the bottom, the manufactured PCB of the virtual PCB example circuit.]

\subsubsection{Design Implementation} For the design implementation of the automated composition of circuit blocks we built a python library we call {
    Swoop
}, that allows the manipulation of Fusion 360 files using python. Using {
 Swoop
} we can create individual Fusion 360 schematic and board files containing the different schematics and PCB blocks information selected by the user, as well as the Mats connection specifications also selected by the user. However, the resulting individual files created with our library, although the connections between blocks are specified, the PCB layout file (board file) remains not routed. For the routing process we use Fusion 360 in the background, which automatically routes and completes missing connections between the different PCB blocks, generating a routed PCB board file. For generating the Gerber files and ERC and DRC checks, we also use Fusion 360 in the background. 

It is worth mentioning that the routing process only needs to connect the missing wires between the PCB blocks, as the PCB blocks are already imported routed following design specifications, making the final PCB design less prone to DRC errors. 

The merging routines described above work as part of the system backend on a server, while Mats and the PCB editor work as the frontend in TypedScheamtics.

\subsection{Extending the circuit blocks library}

The circuit blocks library can be extended by expert users by annotating the syntax language to existing schematic blocks and simply importing both typed schematic blocks and PCB blocks files to TypedScheamtics. Currently, only Fusion 360 schematic and board files can be imported. We also require users to add an STL file of the 3D PCB which is also generated with Fusion 360, and a screenshot of the schematic block. When the files are imported, TypedScheamtics runs syntax validation routines that checks the validity of typed schematic file, as well as ERC and DRC checks. 

\subsection{Extending the language syntax}

The language syntax can be extended by adding more circuit type declarations with their corresponding validation routines. The validation routines are stored as JavaScript code files. Inside the validation routes the declaration types of each hardware protocol bus are also stored. These code files are imported dynamically at runtime as part of the real-time capabilities of Mats, allowing \interfacetype{}s to be edited or added without stopping the program, therefore allowing the language syntax for circuit blocks to scale by providing the possibility for the language syntax to support more protocol interface types. TypedScheamtics currently only supports validations for the following protocol interface types: I2C, SPI, and GPIO.
\section{Design Examples}
\label{sec:examples}

We demonstrate TypedSchematics connection validation futures to guide beginners in merging circuit blocks by showcasing three designs, a Blinky PCB design, a thermostat design, and a Wi-Fi controlled mini-catamaran PCB design.

\subsection{Blinky PCB design}
\label{sec:blinky}

The Blinky PCB design shown in \reffig{fig:teaserfigure} is a minimalist circuit design consisting of an Atmega328p microcontroller, two LEDs, a DC Jack power connector, and a power regulator. The blinky PCB is designed to run a code that simply blink LEDs, this type of circuit is considered the "Hello World" for hardware.

While this PCB design is basic, it shows multiple futures of TypedSchematics. During design, TypedSchematics guides development by advising the designer, through voltage interface matching algorithms, which power regulators are compatible for the Atmega328p microcontroller. For the Atmega328p microcontroller, 5V and 3.3V are available, in this example the 5V regulator was chosen.

When it comes time to connect the LEDs to the microcontroller, the protocol interface matching algorithm guides the designer to select the correct hardware interface to connect the wires between the microcontroller and the LED circuit blocks. In this example, the GPIO hardware interface is the correct interface to connect the LEDs to the microcontroller, TypedScheamtics prevents the connection to any other hardware interface.

The entire PCB design was developed in less than two minutes; By simply adding or removing blocks, new iterations of the PCB design can be generated.

\subsection{Thermostat design and comparison with Fusion 360}
\label{sec:thermostatExample}

A thermostat is a common circuit design used in many commercial electronic devices, including refrigerators, air conditioners, and ovens. Thermostats regulate temperature by activating actuators (e.g., cooling compressors) when the measured temperature has reached a user-entered set point.  In this example, we will discuss how TypedSchematics Mats design aids in the design correctness of the thermostat circuit design. The PCB design experience for this design is omitted in this section in favor of providing a contrast with Fusion 360, as participants use this design for the user study in \refsec{sec:userstudy} where the design, including the PCB, will be further discussed and analyzed.

\reffig{ fig:temperaturecontroller} shows a block-based schematic thermostat design using both TypedSchematics (a) and Fusion 360 (b) to provide a visual comparison. Both tools include the same schematic blocks. As peripherals the design includes: a MCP9808~\cite{MCP9808} temperature sensor, two buttons to set the temperature, an LED indicator that simulates actuators activation, and a display consisting of four 7-segment displays driven by a HTK16K33 IC~\cite{HT16K33}. For computation, an Atmega328~\cite{Atmega328} with an 8~MHz crystal. A DC jack supplies power to the entire design, while a 5V regulator provides the correct voltage to all schematic blocks.

\paperFigureWide[%
    {width=0.80\textwidth},
    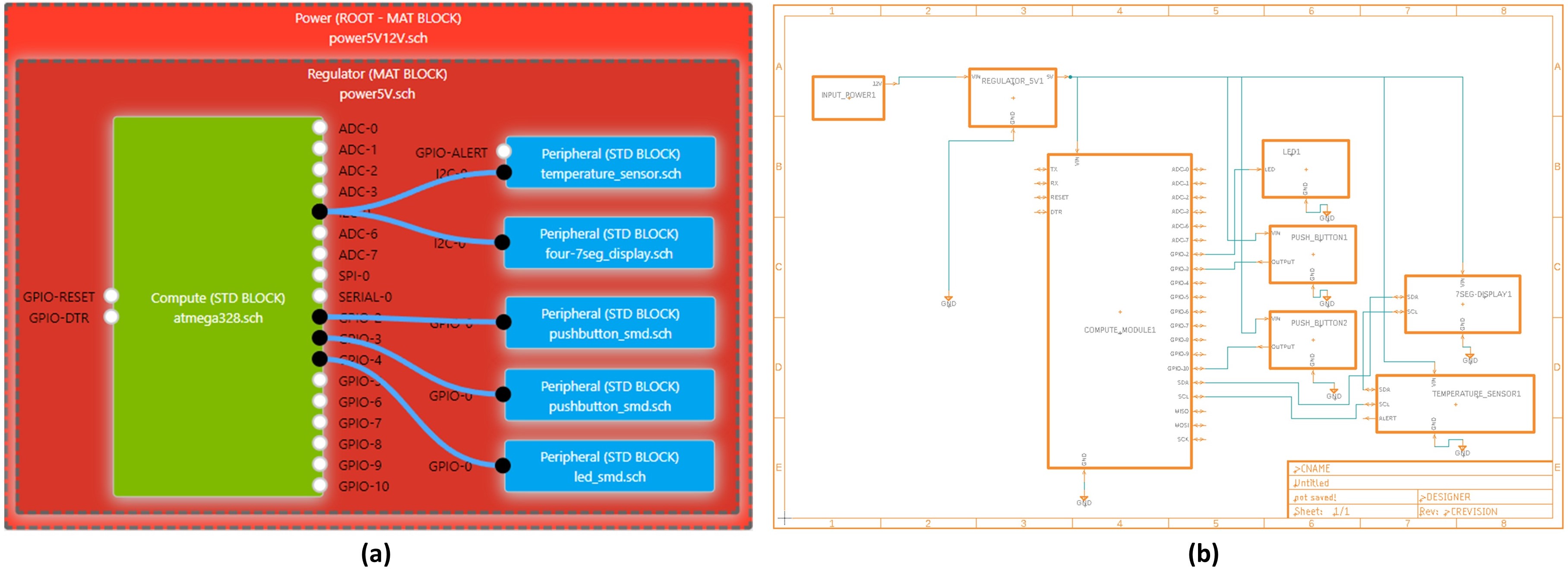,
    {Schematic blocks design of a thermostat schematic using TypedSchematics (a) and Fusion 360 (b). Compared to Fusion 360, TypedSchematics eliminates the need for multiple voltage and ground connections with the use of Mat blocks, reducing interaction costs and wire clutter. TypedSchematics also combines all interface signals into one, further reducing wires and eliminating the need to know how to connect each individual interface signal. When the design is ready, TypedScheamtics warns designers if connections are missing. Fusion 360 (b) design was implemented by a user study participant.},
    fig:temperaturecontroller,
    Two figures, on the left the thermostat design with TypedSchematics and on the right the same thermostat design but with Fusion 360, multiple messy wires are seen in the Fusion 360 design.]

\subsubsection{Circuit blocks nets naming differences}

The thermostat circuit blocks for both TypedScheamtics and Fusion 360 are the same, however the difference relies on the naming of each IO wire. TypedScheamtics circuit blocks use the language syntax described in \refsec{sec:typesyntax} and for Fusion 360 follows a naming conventions commonly used by circuit designers. For regulator blocks we explicitly write the output voltage (e.g. "3.3V"), for microcontroller and sensor circuit blocks whose input voltage can vary, we use "VIN". For the input of the microcontrollers we use the hardware protocol required for each wire (e.g., "GPIO", "SDA", "SCL"). We use alternative names for defined GPIO inputs and outputs  (e.g. "LED", "ALERT"). As an example, the temperature sensor circuit block for Fusion 360 has the following IO wire net names: "VIN", "SDA", "SCL", "ALERT". A button circuit block has: "VIN", "OUTPUT", LED block has: "VIN", "LED" and a 3.3V regulator block has: "VIN", "3.3V". While these community-created naming conventions help guide the user, they are not a standard and IO circuit blocks are often named as desired by the designer.

\subsection{Wi-Fi controlled mini-catamaran PCB design}
\label{sec:catamaran}

The PCB design shown in \reffig{ fig:catamaranpcb} is a more complex design designed for controlling a mini catamaran over Wi-Fi. The mini-catamaran circuit design features an ESP32 microcontroller board with Wi-Fi capabilities, two motor drivers, two buttons and two LEDs, and a DC jack and 5V power regulator as the power supply. 

An interesting aspect of this design is the voltage requirements of the motor drivers, which require the full supply voltage to power the motors, which in this case was provided by a 7.4V lithium battery. By placing the drivers of the motor in the root Mat, outside the voltage regulator Mat, the connection of the motor drivers is made directly to the 7.4V voltage supply.

\paperFigureWide[%
    {width=0.95\textwidth},
    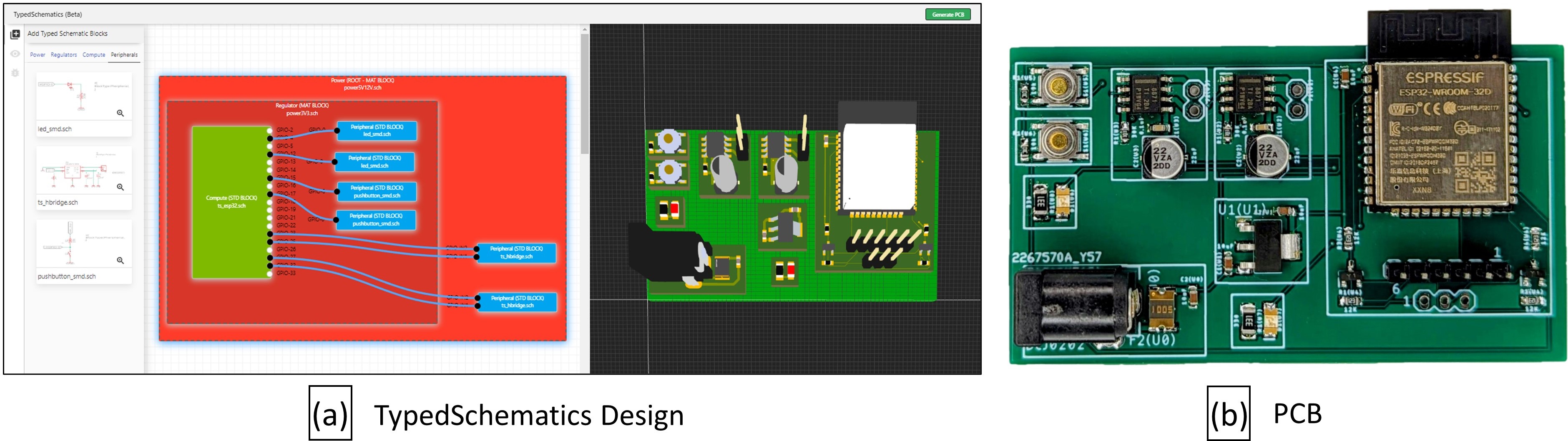,
    {A Wi-Fi controlled mini-catamaran PCB layout designed by high school students using TypedSchematics demonstrates how our tool greatly reduces the skill floor of PCB design. In the figure, (a) shows a screenshot of the design, while (b) shows the assembled PCB design after manufacture.},
    fig:catamaranpcb,
    Two figures, on the left the Wi-Fi controlled mini-catamaran PCB design with TypedSchematics and on the right the assembled PCB design after manufacture.]


The mini-catamaran PCB board was designed by high school students during a STEM summer camp on Unmanned Surface Vehicles (USVs). Before development, the blinky PCB design was used as an example to teach the students how to use TypedSchematics. After learning how to use TypedScheamtics, a process that took around 15 minutes, the students completed the mini-catamaran circuit design without any assistant in around 40 minutes. The design experience of the high-school students is discussed in ~\refsec{sec:results}.

\section{User Study}
\label{sec:userstudy}

To demonstrate how TypedSchematics guides users in the merging process of circuit blocks with the support of real-time validation of connection errors between circuit blocks, we conducted a small user study. In the user study, participants were given the task of designing the thermostat circuit design presented in~\refsec{sec:thermostatExample}, using both TypedSchematics and Fusion 360.

In the user study, TypedSchematics' schematic block reuse method is compared to hierarchical sheets in Fusion 360, while TypedScheamtics' PCB blocks reuse method is compared to modular design blocks, which is similar to device snippets, in Fusion 360.

Our user study demonstrates the capabilities of our tool through a quantitative study, however, it also has a qualitative study, taking into account the opinions and comments of the participants given through a survey at the end of the study. Below, we describe the study. \refsec{sec:results} shows our findings and \refsec{sec:discussion} discusses the results. 

\subsection{Participants}

We recruited seven participants (P01-P7) from our university who recently completed a quadcopter circuit design course using Fusion 360. All participants were third and fourth year university students aged between 18 and 23 years. All recruited participants had intermediate level knowledge of circuit design acquired in the circuit design course. Participants were rewarded with a \$40 Amazon gift card for a two-hour user study.

\subsection{Structure}

The user study consisted of three phases. First, the first half of a Zoom interview where participants were taught how to design a blinky PCB, described in~\refsec{sec:blinky}, using Fusion 360 and TypedScheamtics to close any knowledge gaps. Secondly, the second half of the Zoom interview, where participants where tasked with the design of a thermostat using both TypedSchematics and Fusion 360. Finally, a feedback phase where participants gave their opinions and comments comparing both tools through a questionnaire.

Participants were provided with ten circuit blocks for both Fusion 360 and TypedScheamtics, some not part of the required design. In the case of Fusion 360, participants were emailed a schematics file and board file containing the circuit blocks. For TypedSchematics, participants were sent a web link, different from the one used for tutorial purposes~\footnote{TypedSchematics user study - \url{url-retracted}}, containing the circuit blocks preloaded. For fairness, we designed the Fusion 360 circuit blocks with IO wires names that made sense to identify how to connect the wires (e.g. SDA, SCL, CLK) as mentioned in \refsec{sec:thermostatExample}; Otherwise, it would have been impossible for participants to identify how to connect the circuit blocks provided in a reasonable time using Fusion 360.


During the design of the thermostat participant's computer screen was recorded for measuring the overall design time between both tools. After finishing both designs, they were asked for their opinion on their design experience using both tools and filling a questionnaire, ending the Zoom meeting.

\subsection{Objectives} 

The primary goal of the user study is to evaluate the participant's ability to correctly merge circuit blocks whose design information is unknown, mimicking a scenario in which these circuit blocks are created by the broader community of circuit designers and not by peer developers. Our hypothesis is that due to the lack of connection validations, Fusion 360 designs would have connection errors compared to TypedScheamtics. As a secondary objective we want to compare the design speed between both tools. We hypothesize that TypedScheamtics designs would be faster than Fusion 360 due to reduced interaction costs using Mats. The last objective of the user study is to also evaluate, through a questionnaire, the participants' design support experience with TypedScheamtics in merging circuit blocks by asking participant their confidence in making connections and opinions.

\subsection{Questionnaire} 

To capture the design support experience of participants, participants were asked to complete a questionnaire consisting of 15 questions divided into three sections: schematic blocks merging process, Mats schematic model representation, and general questions. The first two sections of questions centered on understanding the participants' usability experience regarding TypedSchematics compared to Fusion 360. The general questions section, centered on understanding any possible barriers to participants in adopting TypedSchematics for future designs. For each section, participants also had the option to leave personal feedback.

One of the concerns with the Mats schematic model representation was that participants might not fully understand the model and how it works internally. In order to understand whether the model is easy to learn, the questionnaire also included four questions to assess participants' understanding of Mats.

\section{Results}
\label{sec:results}

Results show significant differences in design errors and design time between TypedSchematcs and Fusion 360. Below we show these differences and also include participants' preferences and experience with TypedScheamtics design support.

\subsection{Design errors}

To merge schematic blocks, participants used TypedScheamtics Mats reuse method with real-time connection validations and hierarchical sheets reuse method with Fusion 360.

With TypedSchematics, participants did not present any design mistakes when merging schematic blocks. Despite the help provided in naming wire nets with explicit names to facilitate connections and participants having an intermediate circuit design experience, four of the seven participants had one or more design errors using Fusion 360. All participants with design errors had a power-related connection error, the most common error was using the wrong voltage regulator. Of the four participants, two also had wire connection errors, connecting wires to the wrong hardware interface.

For merging PCB blocks, participants used our 3D editor to merge PCB blocks in TypedSchematics and modular device blocks reuse method, also known as device snippets, in Fusion 360. Participants had no design errors when placing the PCB design blocks using both tools. However, since the wire connections are defined in the schematic design, participants using Fusion 360 ended with connection errors in the PCB layout design as well.

\subsection{Design time}

The results, shown in \reffig{ fig:designtime}, show that, on average, the design time of participants for the schematic design was 3.9 times faster with TypedSchematics than with Fusion 360. Taking the participants on average 16.58 minutes to merge the schematics blocks in Fusion 360 and 4.2 minutes in TypedSchematics.

For the PCB design, design times for both TypedScheamtics and Fusion 360 were similar, an average of 2.5 minutes with TypedScheamtics and 2.9 minutes with Fusion 360, not a significant difference.

\paperFigure[%
    {width=\columnwidth},
    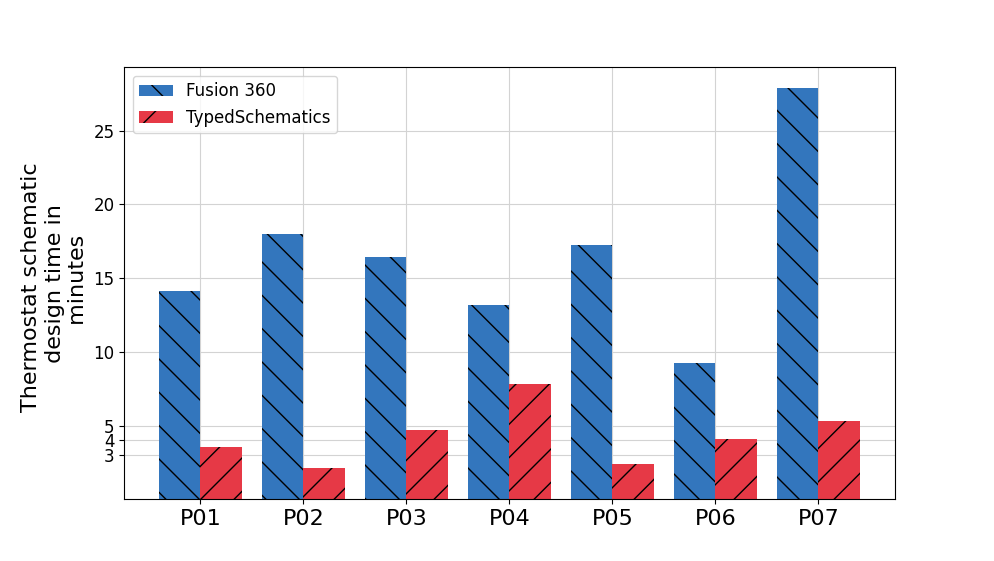,
    {Design time of a thermostat schematic with schematic blocks, described in \reffig{ fig:temperaturecontroller}, using Fusion 360 and TypedSchemtics. On average, participants' design time was 3.9 times faster with TypedSchematics than with Fusion 360.},
    fig:designtime,
    A graph showing on the y-axis the minutes from 0 to 25. On the x-axis, seven participants labeled P01 to P07. For both participants two bars, Fusion 360 in blue and TypedSchematics in red. For all participants, the Fusion 360 bar is larger, indicating more design time.]

\subsection{Design support experience}

The design support experience of participant with TypedSchematics in connecting schematic blocks was assessed through confidence weighted multiple-choice questions. The results in table~\reftab{tab:confidence} show the difference in participants' confidence when making connections using both TypedScheamtics and Fusion 360. Overall, participants felt more confidence connecting schematics blocks using TypedSchematics than Fusion 360. 

\begin{table*}
    \caption{Participants confidence making connections}
    \label{tab:confidence}
    \begin{tabular}{cccccl}
        \toprule
        Question             & Completely Confident & Fairly Confident & Somewhat Confident & Slightly Confident & Not Confident\\
        \midrule
        Signal wire connections \\ using TypedSchematics  & 5 & 2 & 0 & 0 & 0 \\
        \midrule
        Power wire connections \\ using TypedSchematics   & 5 & 2 & 0 & 0 & 0 \\
        \midrule
        Signal wire connections \\ using Fusion 360   & 0 & 3 & 3 & 0 & 2 \\
        \midrule
        Power wire connections \\ using Fusion 360 & 0 & 2 & 3 & 2 & 0 \\
        \bottomrule
    \end{tabular}
\end{table*}

For the PCB layout design we asked participant directly since the PCB design editor was a last minute addition to our tool. All participants preferred TypedScheamtics 3D editor for placing PCB blocks because of the 3D view; however, all participants felt equally confident using both tools.

\subsection{Usability experience of mats and hierarchical sheets}

In terms of usability, six of the participants preferred Mats for merging schematics blocks, and one person preferred the Fusion 360 hierarchical sheets approach. 

One of our concerns was the ease of learning of Mats. Through a four-question questionnaire about the different types of blocks, what connections do mat blocks connect internally (voltage and ground), among other questions, we quickly assessed how well participants understood Mats. Six participants completed the questionnaire without errors, only one participant had comprehension errors on two questions.

Overall, participants felt that using mat blocks helps reduce the visual information overload that occurs with the accumulation of power signals, making the design look cleaner and less cluttered.

\subsection{TypedSchematics adoption}

To conclude the survey, we asked participants a general question about TypedSchematics adoption. All participants responded that one of the barriers that would prevent the adoption of TypedSchematics is the lack of more circuit blocks and the lack of support for other hardware interfaces (e.g., USB, HDMI). Although these interfaces can be added in the syntax language, validations for these interfaces are not yet implemented in our tool. Five participants also mentioned that TypeSchematics is great as a template for quickly getting started with new PCB designs.

\section{Discussion}
\label{sec:discussion}

Through the user study, we made several observations that further show the design differences with TypedSchematics compared to Fusion 360 in merging schematic blocks. In this section, we discuss these observations and also quote some of the user experiences that were gathered through the recordings and questionnaires. We end the section by also discussing the design experience of the high school students who designed and built the Wi-Fi controlled mini-catamaran PCB using TypedScheamtics.




\subsubsection{TypedSchematics vs Fusion 360 design support}

Despite designers had an intermediate knowledge of circuit design with Fusion 360, more than half of the participants made some design mistakes when connecting circuit blocks. This result was expected since Fusion 360 does not have any method that verifies the connection between circuit blocks. The same result is applicable to other traditional PCB design tools (e.g., Altium, EasyEDA, KiCAD), since all of them lack a system for validating connections between circuit blocks. In contrast, TypedScheamtics participants' designs were error-free.

TypedSchematics design supporting in merging schematic blocks is also seen in the considerable reduction in design time, which was 3.9 times faster than with Fusion 360. It was observed that the difference in design time mostly resulted from participants spending great time deciding how to correctly make power and signal connections. In Fusion 360, participants had trouble deciding which voltage regulator to use, either 3.3V or 5V, as there is no information in Fusion 360 that indicates the voltage required for the different circuit blocks such as the microcontroller or temperature sensor blocks where the power wires only indicate "VIN" as voltage net information. Consequently, participants spent a lot of time analyzing the internal diagrams of each schematic block to determine which voltage regulator to use. Despite spending a lot of time figuring out the voltage required for each circuit block, many participants had a power connection design error in Fusion 360. 

In Fusion 360 participants also had spent great time searching for how to connect unknown signals, as is the case with the ALERT signal in the temperature sensor circuit block. For example, P02 spent 1.5 minutes searching online for data sheets and e-blogs to obtain information about the type of ALERT signal interface and its capabilities. The participant later discovered that the connection interface type for the ALERT signal was of type GPIO.

We also observed that participants did not check for I2C address conflicts in the case of connecting both the temperature sensor and the display circuit, since both require an I2C connection. Participants also did not verify master-slave constraints. Both the temperature sensor and the display circuit have a different I2C address; however, if this were not the case, we believe that most participants would have made a design error in Fusion 360. 

In TypedSchematics, participants did not encounter any of these problems as the tool validates voltages, hardware protocol types, and missing connections, guiding participants through the design process by displaying useful error warnings when participants make connection mistakes.

\subsubsection{Participants overall experience}

Overall, with TypedSchematics, participants found the design experience easier than with Fusion 360 for merging circuit blocks. P02 described TypedSchematics design experience as follows:

\begin{displayquote}
"Is not work at all, is very simple to use"
\end{displayquote}

In contrast, P02 also described the design experience with Fusion 360 as follows:

\begin{displayquote}
    "Is not fun. Schematic design, which is a high level design compared to PCB layout, should be much easier".
\end{displayquote}

Mats' schematic representation model was also highly preferred over the traditional one used by Fusion 360. Participants mentioned that it made the design much clearer and easier to understand for connecting different blocks together. In the case of the participant who preferred Fusion 360's schematic model representation over TypedSchematcs, the participant mentioned that the decision was due to the inability of TypedSchematics to manually isolate grounds, an advanced design technique that separates the ground connections into digital and analog ground connections to reduce noise. A future that TypedSchematics does not yet support as it is a PCB design tool intended for beginners.

Overall, participants felt more confident connecting blocks to each other and expressed that TypedSchematics hugely reduces the level of expertise, making design easier. Although our tool was successful in designing beginner level PCBs, a serious concern arose on our part when several participants also expressed job insecurity concerns after using TypedSchematics because they mentioned that PCB design had become "very easy."

\subsubsection{Traditional PCB design tools, Block-based PCB design tools, and TypedScheamtics}

While block-based PCB design tools such as Altium Upverter have already demonstrated forms of connection validations and schematic merging capabilities, and traditional PCB design tools make it possible to share schematic blocks with the circuit designer community. TypedScheamtics is a PCB design tool that addresses the different design challenges from both traditional PCB design tools and block-based PCB design tools with a unique single approach, creating a syntax language that aids in merging, validating and sharing of circuit blocks. 

Compared to Altium Upverter, TypedScheamtics generates fully routed PCB designs and not only schematics, expands connection validations, and allows users to scale the circuit blocks library. Compared with traditional PCB design tools, TypedScheamtics provides better circuit block merging support.

TypedScheamtics also brings a flow-based connection approach between circuit blocks with Mats, making connections between blocks more interactive.

\subsubsection{Reduction of PCB design skill-floor. High school students experience.} 

One of the main concerns we had with high school students was with the ability to understand the basics of how circuits work in order to interact with TypedSchematics. With the push for STEM in secondary education in recent years, we find that students already have a good basic understanding of how circuits work.

Prior to the PCB design, students also created a prototype using breakout boards for the entire design, so the students already had good understanding of the Wi-Fi controlled catamaran circuitry described in \refsec{sec:catamaran}. Circuit designs that were not in TypedScheamtics were simply downloaded from breakout board vendors, typed, and rapidly imported into our tool.

The process for teaching high-school students how TypedScheamtics works was the same as described in the user study for intermediate designers and taking the same time. While we were very skeptical, the PCB design completed by the high-school students took about 40 minutes and, to our surprise, was completed without any help. The final PCB is indistinguishable from a PCB designed by circuit designers with more experience.

Overall, high school students were able to design the mini-catamaran PCB without design errors, even students mentioned that the design was very easy with the connection validations support, a task that would otherwise be difficult with existing tools like Sparkfun ALC or Altium Upverter without the help of expert engineers. The experience with high-school students made us realize the potential of our tool to significantly reduces the PCB design skill-floor.

\section{Applications, Limitations and Future Work}
\label{sec:limitations}

\subsection{Limitations}
Our tool is still a research tool and still has multiple limitations. On the schematic side, some major limitations are the ability to make connections with multiple grounds, the inability to check for current constraints, and the limited number of protocol interfaces and circuit blocks available. While our tool provides mechanisms to extend both protocol interfaces and circuit blocks, the ability to connect multiple grounds and check for current constraints is still a work in progress.

On the PCB side we have multiple limitations. There is a limitation on trace width when routing between circuit blocks, as Fusion 360 manages the routing, and we currently are unable to control the trace widths between blocks. Another limitation is the empty space left in each individual circuit block and the inability to make circuit configurations using circuit components (e.g., selecting an I2C address by changing resistors). Finally, another limitation is that our tool is currently limited to 2 layers.

Another important limitation is that although our tool makes great efforts to detect errors in imported circuit blocks, design errors may go undetected, affecting other circuit design. Other techniques to detect errors should be considered. Additionally, currently the circuit block import mechanism is done through a public server folder, a graphical interface has not yet been implemented.

While there are multiple limitations, we believe that our tool is capable of supporting beginners in low-end PCB designs.

\subsection{Applications and future work}

We see our circuit reuse approach having application in three different ways. First, integrated into existing traditional PCB design tools to improve circuit block reuse. Secondly, as a stand-alone block-based PCB design tool, as is our current focus. Third, and part of our future work, integrated into the breakout boards design process to provide a system that bridges circuit prototyping with PCB design.

\section{Conclusion}
\label{sec:conclude}

Block-based PCB design tools (e.g., Sparkfun ALC, Upverter Modular) have gained interest for rapidly creating circuit designs from reusable circuit blocks. While some of these tools implement connection validations and support schematic composition, they are unable to generate complete PCB designs and lack circuit blocks library scalability by users. Block-based PCB design tools often rely on traditional PCB design tools (e.g., Fusion 360) for composing circuit blocks, which in turn present design challenges when merging circuit blocks. These challenges include: correctly connecting hardware protocol interfaces, supplying the correct voltages to the schematic blocks, detecting if some signals on the schematic blocks are optional. 

TypedScheamtics addresses these design challenges from both traditional PCB design tools and block-based PCB design tools with a unique single approach, creating a syntax language that aids in merging, validating and sharing of circuit blocks. TypedSchematics provides interactive feedback on potential design problems including mismatch voltages, mis-connected signals in complex interfaces, and missing (but necessary) connections. Our tools automated composition of circuit blocks generates fully routed PCB designs. A standardized syntax language allows our circuit block library to be user-scalable. 

TypedSchematics's Mats abstraction eliminates the need for repetitive voltage and ground connections, reducing interaction costs. A flow-based programming paradigm is seamlessly integrated into Mats, making it possible to perform \interfacecheking{} in real time.

We demonstrate TypedSchematics capabilities with three PCB design examples, one designed by high-school students. We further demonstrate the design challenges of existing PCB design tools and how TypedSchematic addresses these challenges through a user study. Results show that when designing a thermostat PCB using circuit blocks, more than half of the participants had a design error with Fusion 360 and none with TypedScheamtics. In designing the thermostat PCB, participants were 3.9 times faster with TypedSchematics than with Fusion 360. Feedback from participants via a questionnaire also showed that participants felt completely confident in making connections with TypedSchematics compared to Fusion 360, where they felt somewhat confident. Ultimately, we hope our work better supports designers in merging circuit blocks and extends the reach of PCB design to beginners for building custom, personalized circuits.

\begin{acks}
  Acknowledgements
\end{acks}

\bibliography{paper.bib}
\bibliographystyle{ACM-Reference-Format}

\end{document}